\documentclass[]{aa}  

\renewcommand{\linenumbers}{}
\renewcommand{\nolinenumbers}{}
\renewcommand{\runninglinenumbers}{}

\usepackage{graphicx}
\usepackage{txfonts}
\usepackage{xcolor}
\usepackage{booktabs}
\usepackage{orcidlink}
\usepackage{xparse}
\usepackage{enumitem}
\usepackage{amsmath}
\usepackage{makecell}
\usepackage{pifont}
\usepackage{xspace}
\usepackage{siunitx}

\usepackage{tikz}
\usepackage{subcaption}         
                                
\usepackage{lscape}            
                               
\usepackage{placeins}

\usepackage{natbib,twoopt}

\bibpunct{(}{)}{;}{a}{}{,}             
\makeatletter
  \newcommandtwoopt{\citeads}[3][][]{\href{http://adsabs.harvard.edu/abs/#3}
    {\def\hyper@linkstart##1##2{}
     \let\hyper@linkend\@empty\citealp[#1][#2]{#3}}}
  \newcommandtwoopt{\citepads}[3][][]{\href{http://adsabs.harvard.edu/abs/#3}
    {\def\hyper@linkstart##1##2{}
     \let\hyper@linkend\@empty\citep[#1][#2]{#3}}}
  \newcommandtwoopt{\citetads}[3][][]{\href{http://adsabs.harvard.edu/abs/#3}
    {\def\hyper@linkstart##1##2{}
     \let\hyper@linkend\@empty\citet[#1][#2]{#3}}}
  \newcommandtwoopt{\citeyearads}[3][][]
    {\href{http://adsabs.harvard.edu/abs/#3}
    {\def\hyper@linkstart##1##2{}
     \let\hyper@linkend\@empty\citeyear[#1][#2]{#3}}}
\makeatother

\begin{document}

   \title{The star formation history of NGC 2276: Comparison between SED modeling and hydrodynamical simulations}

   \author{L. Matijević
          \inst{1,2}\orcidlink{0009-0004-2049-7701}
          \and
          A. Marasco\inst{2}\orcidlink{0000-0002-5655-6054}
          \and
          N. Tomičić\inst{1}\thanks{Corresponding author: ntomicic.phy@pmf.hr}\orcidlink{0000-0002-8238-9210}          
          \and
          R. Smith\inst{3,4}\orcidlink{0000-0001-5303-6830}
          \and
          P. J. Watson \inst{2}\orcidlink{https://orcid.org/0000-0003-3108-0624}
          \and
          A. Ignesti\inst{5}\orcidlink{0000-0003-1581-0092}
          \and
          I. D. Roberts\inst{6,7}\orcidlink{0000-0002-0692-0911}
          \and
          P. Sell\inst{8}\orcidlink{0000-0003-1771-5531}
          \and
          A. E. Lassen\inst{2}\orcidlink{0000-0003-3575-8316}
          \and
          A. Wolter\inst{9}\orcidlink{https://orcid.org/0000-0001-5840-9835}
          \and
          K. Anastasopoulou\inst{10}\orcidlink{0000-0002-3240-6609}
          \and
          A. Zezas\inst{11,12}\orcidlink{0000-0001-8952-676X}
          \and
          P. Kotoulas\inst{13}
          }

   \institute{Department of Physics, Faculty of Science, University of Zagreb, Bijenička Cesta 32, 10000 Zagreb, Croatia
         \and
         INAF–Padova Astronomical Observatory, Vicolo dell’Osservatorio 5, 35122 Padova, Italy
         \and
         Departamento de Fisica, Universidad Tecnica Federico Santa Maria, Avenida España 1680, Valparaíso, Chile
         \and
         Millenium Nucleus for Galaxies (MINGAL), Valparaíso, Chile
         \and
         Astronomical Institute of the Czech Academy of Sciences, Bo\v cn\'i II 1401, 14100 Prague, Czech Republic
         \and
         Waterloo Centre for Astrophysics, University of Waterloo, 200 University Avenue West, Waterloo, ON, N2L 3G1, Canada
         \and
         Department of Physics \& Astronomy, University of Waterloo, 200 University Avenue West, Waterloo, ON, N2L 3G1, Canada
         \and
         Georgia Institute of Technology, North Avenue, Atlanta, GA 30332, USA
         \and
         INAF–Osservatorio Astronomico di Brera, Via Brera, 28, 20121 Milano, Italy
         \and
         Istituto Nazionale di Astrofisica (INAF) – Osservatorio Astronomico di Palermo, Piazza del Parlamento 1, 90134 Palermo, Italy
         \and
         Physics Department \& Institute of Theoretical and Computational Physics, University of Crete, GR 71003 Heraklion, Greece
         \and
         Institute of Astrophysics, Foundation for Research and Technology-Hellas, GR 71110 Heraklion, Greece
         \and
         Department of Physics, University of Florida, Gainesville, FL 32611-8440, USA
             }
 
  \abstract
  {Analyzing environmental effects in galaxy groups is pivotal for understanding how galaxies evolve in these moderate-density settings. The degree to which ram pressure versus tidal interactions drive the unusual morphologies and kinematics seen in group galaxies remains a subject of active debate.}
  {This study focuses on the nearby galaxy NGC 2276, which has the longest radio continuum tail in galactic groups. It is a member of the NGC 2300 group that possibly experiences both processes, and we aim to determine which has the greatest impact on its overall structure.}
  {We combined broadband images with synthetic narrow band filters around emission line maps from integral field spectrograph to construct spatially resolved spectral energy distributions (SED), which we modeled using the BAGPIPES software package to reconstruct kpc-scale star formation histories. These are compared with those derived from adaptive mesh refinement wind tunnel simulations of an NGC 2276-like system.}
  {We show that the spatial distribution of the oldest stellar populations ($>1.1$ Gyr) is highly symmetric compared to the youngest ones. This is the most compelling evidence so far pointing towards ram pressure being the only morphological disturber of this system, as it primarily affects the gaseous component. Simulated data show similar results, with older stellar populations being more symmetric. Our RPS-only model showed no significant differences in the morphology of the stellar populations compared to the RPS+tidal model with initial separation of 50 kpc.}
  {}


   \titlerunning{NGC 2276 Star Formation History}

   \maketitle

\section{Introduction}
\label{sect:intro}

Group and cluster galaxies have been extensively studied in order to assess their interplay with the host environment \citep[e.g.][]{davis+97, zabludoff+98, koopmann+2004, boselli+2006, roberts+21}. 
Multiple observational \citep[e.g.,][]{Alonso+2006, Park+2007, Boselli+2014, Poggianti+2017, Gordon+2018, Zabel+2019} and theoretical \citep[e.g.,][]{Tonnesen+2007, Kawata+2008, Henden+2018, Tremmel+2019, Vollmer+2021} studies were performed to quantify the effects of gravitational (e.g., tidal interaction) and hydrodynamical (e.g., ram pressure stripping; RPS) interactions between galaxies and their host environment across cosmic epochs.

Considering that in dense environments (massive groups and clusters, n$_\mathrm{e}\geq10^{-3}$ cm$^{-3}$), multiple environmental effects can influence galactic properties at the same time \citep[e.g.,][]{White+91, Kantharia+2005, lin+2023}, it is sometimes challenging to distinguish between them, especially with insufficient resolution and sensitivity. 
Combining spatially resolved studies of the physical properties of galaxies with multiwavelength spectral energy distribution (SED) fitting techniques and dedicated numerical simulations might provide adequate details for this particular problem \citep[e.g.,][]{fritz+2017, Vulcani+21, ding+2024, ditrani+24, watson+25}.

In this work, we study NGC 2276, a lopsided spiral galaxy within the NGC 2300 group \citep[D$_\mathrm{L}$ = 28.5 Mpc,][]{fadda+23}. 
The peculiarity of this group lies in the fact that it is the first galactic group in which diffuse X-ray emission was detected emanating from its intragroup medium \citep[IGM;][]{Mulchaey+93}. 
This group has only 8 confirmed candidate members according to the study of \citet{fadda+23}, with the early-type NGC 2300 (M$_*\simeq10^{11}\,\mathrm{M_\odot}$), and late-type spiral NGC 2276 (M$_*\simeq4\cdot10^{10}\,\mathrm{M_\odot}$) being the most massive ones\footnote{Stellar masses of the two galaxies were estimated using  Spitzer's IRAC 1 and IRAC 2 maps with Eqn. 8 from Querejeta et al. (2015)}. 
It is a rather small group with the extent of its diffuse X-ray emission of $\sim0.33$ Mpc \citep{davis+96}, and its $\mathrm{R_{vir} \approx 0.8 \, Mpc}$\footnote{Assuming $\mathrm{R_{vir}} = 1.85 \left( \mathrm{T/10 \, keV} \right)^{0.5} \left( 1+z \right)^{-1.5} \mathrm{h_{100}^{-1} \, Mpc}$ \citep{mulchaey+2000}}.
Furthermore, this galactic group has also been extensively studied in other wavebands \citep[e.g.,][]{Gruendl+93, davis+97, tomicic18, roberts+24}.

The late-type spiral NGC 2276 exhibits an intricate lopsided morphology. 
Prominent features of this asymmetry include extended X-ray emission \citep{rasmussen+2006} and a 100-kpc long low-frequency radio continuum tail \citep{roberts+24} on the galaxy's trailing eastern side.
Radial velocity difference \citep[$\sim 400$ km/s,][]{fadda+23} between NGC 2276, and its nearest neighbor NGC 2300 is substantially smaller than the transverse velocity obtained from plasma-ageing \citep[$\sim 870$ km/s,][]{roberts+24}.
Thus, NGC 2276 moves through the IGM of an estimated density of $\sim5\times10^{-27} \mathrm{\,g\,cm^{-3}}$ \citep{davis+96} and a temperature of T$\sim10^7$ K \citep{Mulchaey+93, rasmussen+2006}, with the total 3D velocity of $\sim970$ km/s \citep{roberts+24} in a mostly westward direction on the plane of the sky, subjecting it to significant ram pressure.
Consequently, the leading western edge experiences a compression in the ISM which can be observed at several wavelengths \citep[e.g.,][]{Gruendl+93, davis+97}, characterized by strong, asymmetric H$\alpha$ emission \citep{Gruendl+93, tomicic18}.
These conditions drive intense star formation yielding a star formation rate (SFR) of $12\pm4 \mathrm{\,M_\odot\,yr^{-1}}$ and an unusual $\sim2-3$ dex variation in the star formation efficiency (SFE) of the bulk of molecular gas \citep{tomicic18}. The value of the SFR is adapted from \citet{tomicic18} and rescaled from their assumed distance of $\sim 35$ Mpc to the updated NGC 2300 group distance of $28.5$ Mpc established by \citet{fadda+23}.
Additionally, numerous compact X-ray sources \citep[e.g.,][]{wolter+15, Anastasopoulou+19}, including a potential candidate for an off-nuclear intermediate-mass black hole \citep[IMBH,][]{mezcua+15} were discovered in this galaxy.
While the ram pressure scenario provides a strong explanation for the extended tail and compressed leading edge, truncation and morphological asymmetry can also be signatures of tidal interactions \citep[e.g.,][]{Barnes+92}.
This possibility is suggested by the small projected separation of $\sim52$ kpc between NGC 2276 and NGC 2300.
The possibility of a tidal interaction was previously investigated through simulations by \citet{wolter+15}, where the authors derived that the tidal interaction had a marginal effect.

We have studied this galaxy in our previous paper \citep[][hereinafter M26]{Matijevic+26} to provide a preliminary investigation on the predominant effect. 
To show this, we analyzed the broadband Hubble Space Telescope's (HST) Wide Field Camera 3 (WFC3) images and combined them with Spitzer's Infrared Array Camera (IRAC) images which we both smoothed down to 2.7$''$ ($\sim0.4$ kpc). 
We found that both the HST NUV and the IRAC 1 and IRAC 2 NIR images (proxies for the young and the old stellar populations, respectively) showed a lopsided morphology, but with the former showing signs of a truncation compared to the latter.
Because RPS affects primarily gaseous component, and in turn young stars, while the tidal interaction affects the distribution of all stellar populations, our finding indicated that both ram pressure and gravitational interactions might be at work.
We have additionally traced that there is visible spiral arm unwinding happening in this galaxy that could potentially be connected to ram pressure \citep[e.g.,][]{bellhouse+21, lassen+26b}.

The main limitation of our previous work was in the assumption that UV and NIR light are perfect tracers for the young and the old stellar populations, respectively.
Due to this assumption, we were not able to fully separate RPS and tidal interaction in this galaxy, which may be a prototypical case of both effects working together in a galactic group.
In spite of this, ram pressure being the predominant force would be an unusual occurrence given the low number of group members (thus reducing the probability of strong RPS, as the number of members has a correlation with both density and mass of the IGM) and the proximity of two galaxies (which raises the likelihood that tidal force will have a dominant role).
In the present study we go beyond our initial assumption and use spatially resolved SED modeling technique to infer the detailed star formation history (SFH) and spatial distribution of the stellar population of different ages, and compare the results with predictions from dedicated numerical simulation of a NGC 2276-like system.
Our conclusions will be based on the spatial distribution of the oldest stellar population, and on the analysis of the SFH maps.
In this work we focus only on the SFH of this system. Other ancillary results will be shown either in the appendix.
We stress that \citet{fadda+23} already investigated SED fitting for this galaxy. However, their work focused on individual aperture regions selected in various parts of the galactic disk. Here we significantly expand their analysis by focusing on the entire galactic disk, and using much smaller resolution elements.

This work is structured as follows.
In Sect. \ref{sect:data} we describe the reduction of the dataset used in this analysis. 
In Sect. \ref{sect:results} we show the result of our analysis of SFH in NGC 2276.
In Sect. \ref{sect:sim} we present the details of the dedicated simulation of RPS in NGC 2276 and how it affects its SFR. 
In Sect. \ref{sect:discussion} we provide a physical interpretation of our results. 
Sect. \ref{sect:summary} summarizes our work.  
Throughout this work, we assume a standard $\Lambda$-CDM cosmology framework with $\Omega_m$ = 0.3, $\Omega_\Lambda$ = 0.7, and $H_0$ = 70 km s$^{-1}$ Mpc$^{-1}$ \citep{planck20}, as well as the initial mass function (IMF) by \citep{kroupa+2001}.

\section{Observations and data reduction}
\label{sect:data}

\begin{figure*}[h!]
    \centering
    \includegraphics[width = 0.99\hsize]{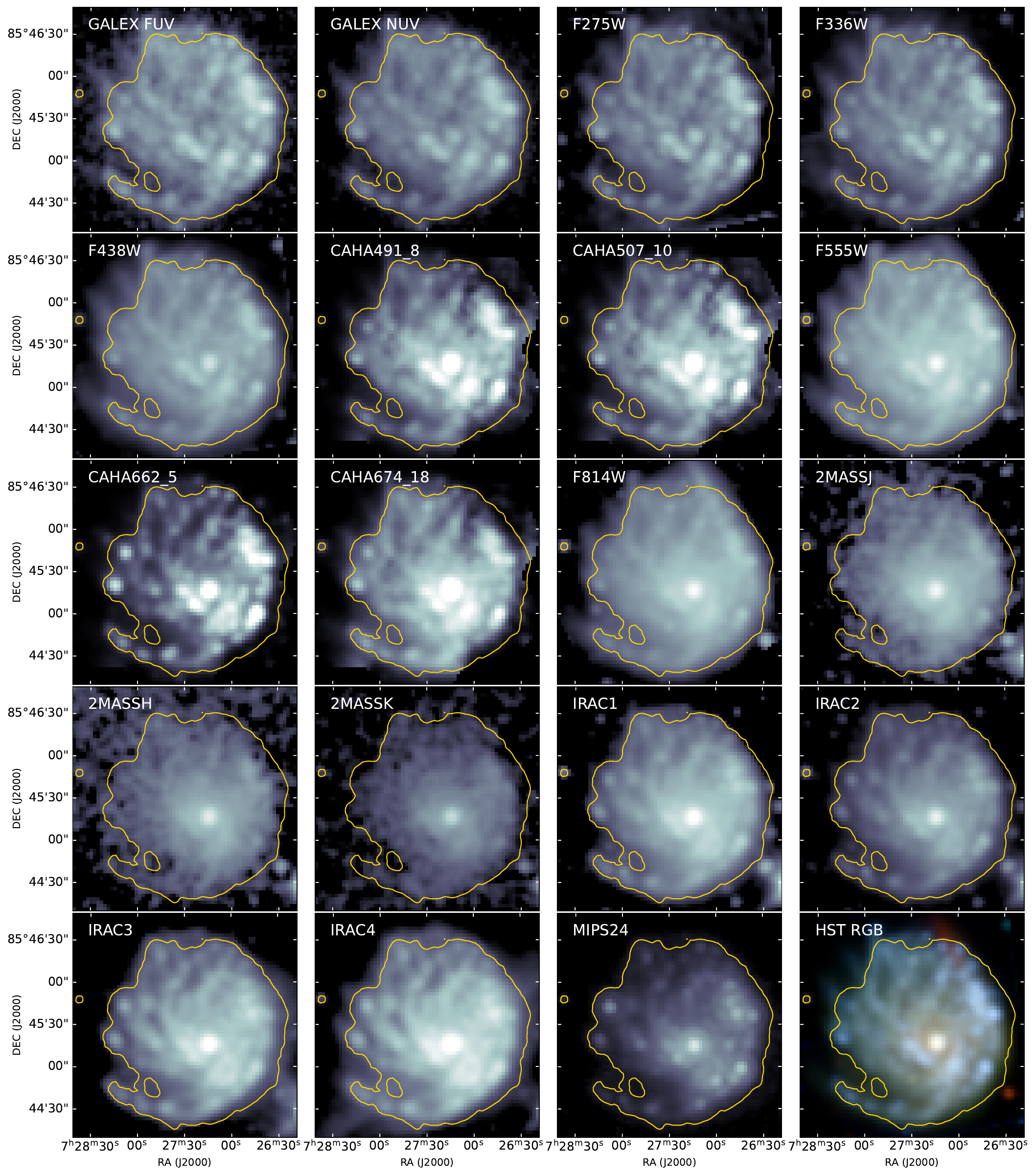}
    \caption{Images of NGC 2276 in different filters used in this work. Bottom-right panel shows the RGB composite image from M26. All images are smoothed at a common resolution of $5.9^{\prime\prime}$, and have pixel scale of $2.4^{\prime\prime}$. 3-$\sigma$ level contour from the MIPS24 map is plotted on top of each map. Top-side of the panels is north, and the left-side is east.
    }
    \label{fig:filters}
\end{figure*} 

In this work, we constrain the spatial distribution of stellar populations of different ages using a collection of publicly available multi-wavelength images and proprietary data. Table \ref{tab:filters} provides a list of the available filters and their main properties, and in Fig. \ref{fig:filters} we show how NGC 2276 looks in these different filters. Below we describe how we reduced the data and made them suitable for SED fitting.

\begin{table}[t!]
\caption{General characteristics of instrument filters used in this work.}
\label{tab:filters}
\centering
\begin{tabular}{lllll}
\hline                                                                                    
\hline                                                                                    
Filter name & $\theta$ [$''$] & $\lambda_\mathrm{eff}$ [$\mu$m] & $\Delta\lambda$ [$\mu$m] & Exp. time \\
\hline                                                                                    
Galex FUV       & 4.0                 & 0.155                  & 0.027            & 2015s      \\
Galex NUV       & 5.6                 & 0.230                  & 0.077            & 2146s      \\
HST F275W       & 0.08                & 0.272                  & 0.042            & 4x700s     \\
HST F336W       & 0.07                & 0.336                  & 0.051            & 4x700s     \\
HST F438W       & 0.07                & 0.432                  & 0.059            & 4x700s     \\
CAHA491\_8      & 2.7                 & 0.491                  & 0.008            & 3x900s     \\
CAHA507\_10     & 2.7                 & 0.506                  & 0.011            & 3x900s     \\
HST F555W       & 0.07                & 0.524                  & 0.159            & 4x700s     \\
CAHA662\_5      & 2.7                 & 0.662                  & 0.005            & 3x900s     \\
CAHA674\_18     & 2.7                 & 0.674                  & 0.018            & 3x900s     \\
HST F814W       & 0.08                & 0.796                  & 0.177            & 4x700s     \\
2MASS J         & 3.1                 & 1.229                  & 0.162            & 1419s      \\
2MASS H         & 3.1                 & 1.639                  & 0.251            & 1419s      \\
2MASS K         & 3.1                 & 2.152                  & 0.262            & 1419s      \\
IRAC 1          & 1.66                & 3.507                  & 0.684            & 712s       \\
IRAC 2          & 1.72                & 4.437                  & 0.865            & 712s       \\
IRAC 3          & 1.88                & 5.628                  & 1.391            & 712s       \\
IRAC 4          & 1.98                & 7.589                  & 2.529            & 712s       \\
MIPS 24         & 5.9                 & 23.209                 & 5.296            & 712s      \\
\hline                                                                                   
\end{tabular}
\tablefoot{$\lambda_\mathrm{eff}$ is the effective wavelength of a filter, $\Delta\lambda$ is the passband rectangular width, and $\theta$ is the full width at half maximum of the Gaussian point spread function for the corresponding filter. } 
\end{table}

Images in the far- and near-ultraviolet (FUV and NUV) were taken from three individual pointings of the Galaxy Evolution Explorer \citep[GALEX;][]{galex}. 
Two of them were observed by the All Sky Imaging Surveys (AIS, obs. IDs 6387804283080802304 and 6387804288449511424) and one was observed with a Guest Investigator (GI) program (P.I. Jennifer Donovan, obs. ID 3124339393641316352) and were retrieved from the Mikulski Archive for Space Telescopes (MAST). 
They were background subtracted using dedicated background sky maps retrieved from MAST, converted from counts per second (CPS) to surface brightness units, and then stacked into a mosaic using \textsc{Astropy's} \citep{astropy} function \texttt{reproject\_and\_coadd} for which we adopted the parameters $match\_background$ = True and $combine\_function$ = median. 

Data retrieval and reduction of Hubble Space Telescope Wide Field Camera 3 (WFC3) filters F275W, F336W, F438W, F555W, and F814W were described in M26. 

\begin{table*}[t!]
\caption{Parameters adopted for the \textsc{bagpipes} modelling. Numbers in round brackets indicate the interval adopted for the (uniform) priors.}
\label{tab:bp_priors}
\centering
\begin{tabular}{llll}
\hline
\hline
Parameter                                            & Description                                                          &                         & Input values \\
\hline
                                                     &                                                                      & \texttt{dust}           &              \\
\texttt{type}                                        & Attenuation law                                                      &                         & CF00         \\
\texttt{Av}                                          & Absolute stellar attenuation in the                                  &                         & (0.0, 3.0)   \\
                                                     & V band                                                               &                         &              \\
\texttt{eta}                                         & Multiplicative factor on A$_\mathrm{v}$ for stars                    &                         & 3            \\
                                                     & in birth clouds                                                      &                         &              \\
\texttt{n}                                           & Power-law slope of attenuation law                                   &                         & 0.7          \\
\hline
                                                     &                                                                      & \texttt{nebular}        &              \\
\texttt{logU}                                        & $\log_{10}$ of the ionization parameter                              &                         & (-7.0, -1.0) \\
\hline
                                                     &                                                                      & \texttt{SFH model}      &              \\
\texttt{type}                                        &                                                                      &                         & continuity   \\
\texttt{massformed} [$\log_{10} \mathrm{M_\odot}$]   & Total stellar mass formed                                            &                         & (5.0, 10.0)  \\
\texttt{metallicity} [Z$_\odot$]                     &                                                                      &                         & (0.0, 3.0)   \\
\texttt{metallicity\_prior\_mu} [Z$_\odot$]          &                                                                      &                         & 1.0          \\
\texttt{metallicity\_prior\_sigma} [Z$_\odot$]       &                                                                      &                         & 0.5          \\
\texttt{bin\_edges} [Myr]                            &                                                                      &                         & [0.0,  30.0,  100.0,  330.0, \\                                            &                                                                      &                         & 1100.0  3600.0, 13200.0] \\
\texttt{dsfr}                                        & Logarithmic ratio of SFR between                                     &                         & (-5.0, 5.0) \\                                                          & adjacent age bins                                                    &                                         \\
\texttt{dsfr\_prior}                                 &                                                                      &                         & student\_t   \\
\hline
                                                     &                                                                      & \texttt{other}          &              \\
\texttt{t\_bc} [Gyr]                                 & Max age of birth clouds                                              &                         & 0.01         \\
\texttt{redshift}                                    &                                                                      &                         & 0.0079       \\
\hline
\end{tabular}
\end{table*}

In order to better constrain the recent ($t\lesssim30$ Myr) star formation history, we use existing optical integral field unit (IFU) data from the Centro Astronómico Hispano-Alemán (CAHA) facility to generate synthetic narrow-band photometric measurements, which are subsequently integrated into our SED fitting. 
Specifically, these synthetic narrow-band filter maps are generated by making a weighted average over the IFU spectra using the dedicated filter profiles\footnote{The synthetic photometry was measured using \textsc{pyphot} \citep{pyphot} and dedicated filter curves available at the SVO Filter Profile Service at \url{https://svo2.cab.inta-csic.es/svo/theory/fps3/}.}. 
Details on the data calibration of the CAHA IFU are given in Appendix \ref{sect:app-ifu}.

The CAHA filters used in our analysis are $50-180\,\AA$ wide, and cover spectral range with the following lines: CAHA491\_8 with H$\beta$, CAHA507\_10 with [\ion{O}{III}]$\lambda\mathrm{5007}$, CAHA662\_5 with H$\alpha+$[\ion{N}{II}], and CAHA674\_18 with [\ion{S}{II}]$\lambda\lambda\mathrm{6716,31}$.  
We noticed that the total flux in the synthetic F555W filter made by \textsc{pyphot} is lower by $\approx$ 15\% compared to the HST F555W, which is why we multiplied the synthetic CAHA filter maps by 1.15. The wavelength range of the other HST filters is outside of the wavelength coverage of the Calar Alto IFU data which is why we decided to use a single value of the flux correction for the CAHA filter maps.

The Two Micron All Sky Survey \citep[2MASS;][]{2mass} maps were retrieved from NASA/IPAC’s Infrared Science Archive (IRSA) by searching for all individual pointings that cover NGC 2276. 
For each filter 4 pointings were found, which were converted to surface brightness units and stacked into a mosaic using the same function and parameters as for the GALEX maps. 
Afterwards, they were background subtracted by measuring the median background level computed using different circular apertures (r = 6$''$) outside of the galactic disk.

Spitzer's Infrared Array Camera \citep[IRAC;][]{irac} maps were retrieved and reduced as described by M26. 
Spitzer's Multiband Imaging Photometer \citep[MIPS;][]{mips} mid-infrared (MIR) 24 $\mu$m map was retrieved from NASA/IPAC’s IRSA and background subtracted using the same method as for 2MASS maps.

To prepare all our maps for SED fitting, we degraded them to the same resolution and pixel scale of the MIPS 24 map (5.9$''$ angular resolution, 2.45$''$ pixel scale). For smoothing we used \textsc{Astropy's} \citep{astropy} function \texttt{convolve\_fft} with the \texttt{Gaussian2DKernel}, and for reprojecting we used \texttt{reproject\_interp} function.
Before fitting, we corrected our maps for Milky Way dust extinction using the \textsc{Interstellar Dust Extinction} Python package \citep{Gordon2024} with the \cite{gordon+23} dust extinction model with $\mathrm{R(V)}=3.1$, and $\mathrm{E(B-V)}=0.0859$ which is derived from dust maps by \cite{schlafly+11} at the coordinates of NGC 2276.

\section{Analysis of photometric data}
\label{sect:results}

\begin{figure*}[h]
    \centering
    \begin{tikzpicture}
        
        \node[anchor=south west, inner sep=0pt] (base_img) at (0,0) {
            \includegraphics[width=\textwidth]{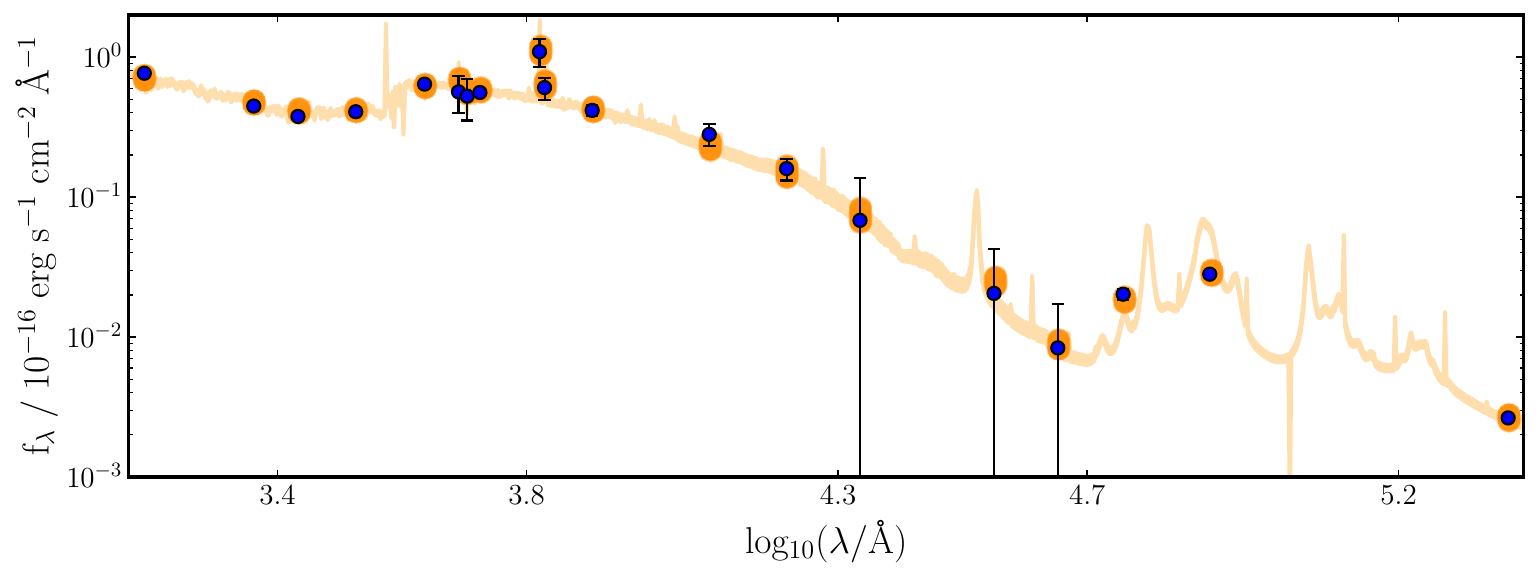}
        };

        \node[anchor=south west, inner sep=0pt] at (1.7cm,1.3cm) {
            \includegraphics[width=3.9cm]{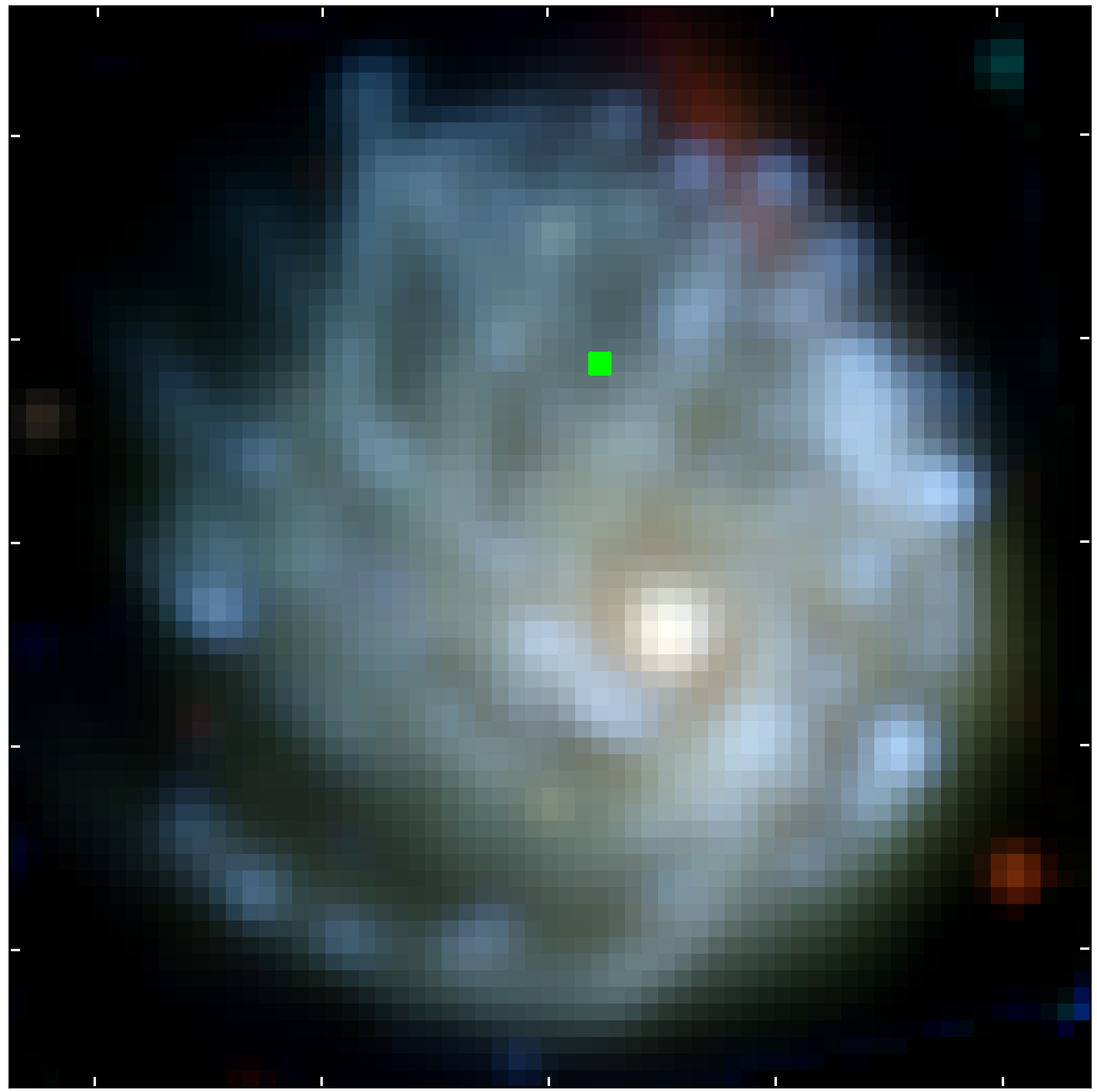}
        };

    \end{tikzpicture}
    \caption{Example of \textsc{bagpipes} fitted SED for a selected spatial bin in the inter-arm region of NGC 2276, shown with green square on the RGB image. Photometric data are shown in blue (the bands used for fitting are shown in Table \ref{fig:filters}). The 16–84th percentile range for the posterior spectrum and the corresponding synthetic photometry points are shown in light and dark orange, respectively. Corner plots for this fit are shown in the Appendix on Fig. \ref{fig:fit_ex1}}
    \label{fig:fit_ex}
\end{figure*} 

In Fig. \ref{fig:filters} we can see that there is a clear east-west asymmetry in NGC 2276 across all of our 19 wavelength bands from FUV to MIR, which was shown in other works as well \citep[e.g.,][]{Gruendl+93, rasmussen+2006, wolter+15}.

To investigate how the environment affected the star formation history (SFH) of this galaxy, and to constrain the timescales at which multiple disturbing mechanisms might have acted on NGC 2276, we used the SED fitting code \textsc{bagpipes} \citep[Bayesian Analysis of Galaxies for Physical Inference and Parameter EStimation,][]{bagpipes}. 
We decided to apply SED fitting for four different configurations: 1) integrated SED measured within a fixed region delimited by an aperture of r$\sim1'$, delimited by the 3-$\sigma$ contour of the MIPS 24 map (see Fig. \ref{fig:filters}), 2) radially binned data, 3) radially binned data divided into eight angular sectors, 4) spatially resolved (with spaxel size of $\sim0.4$ kpc). 
For our radial bins, we use concentric ellipses \citep[assuming an inclination of $\sim20$°;][]{tomicic18} by adopting a minimum separation between bins of 6$''$ ($\sim0.8$ kpc), and then gradually increase it inside-out, up to 20$''$ in the last bin until we achieved a signal-to-noise (S/N) ratio within the annulus considered larger than 5 in all of our maps.
Our binning scheme for both the concentric ellipses and the sectors is shown in the Appendix on Fig. \ref{fig:binning}.
The motivation behind using differently resolved data for SED fitting is to show how resolution can affect the final results, so that the accuracy of galaxy properties retrieved using unresolved photometric data can be carefully assessed.

For all configurations, our unified multiwavelength pipeline fits 19 bands spanning from the FUV to the MIR. 
While direct full-spectrum fitting of the IFU data would preserve fine, age-sensitive spectral indicators (such as stellar absorption lines and the 4000 $\AA$ break), fitting the entire 3D data cube pixel-by-pixel alongside widely separated imaging data is computationally highly demanding for a resolved disk of this scale. 
By converting the key emission-line regions of the IFU data into synthetic narrow-band photometry instead, we seamlessly incorporate these localized star-formation constraints while keeping the pipeline computationally feasible. 
Crucially, test runs excluding these synthetic CAHA filters yielded highly consistent star formation histories, confirming that our hybrid approach robustly captures the underlying stellar populations.

\begin{figure*}[ht!]
    \centering
    \includegraphics[width = 0.99\hsize]{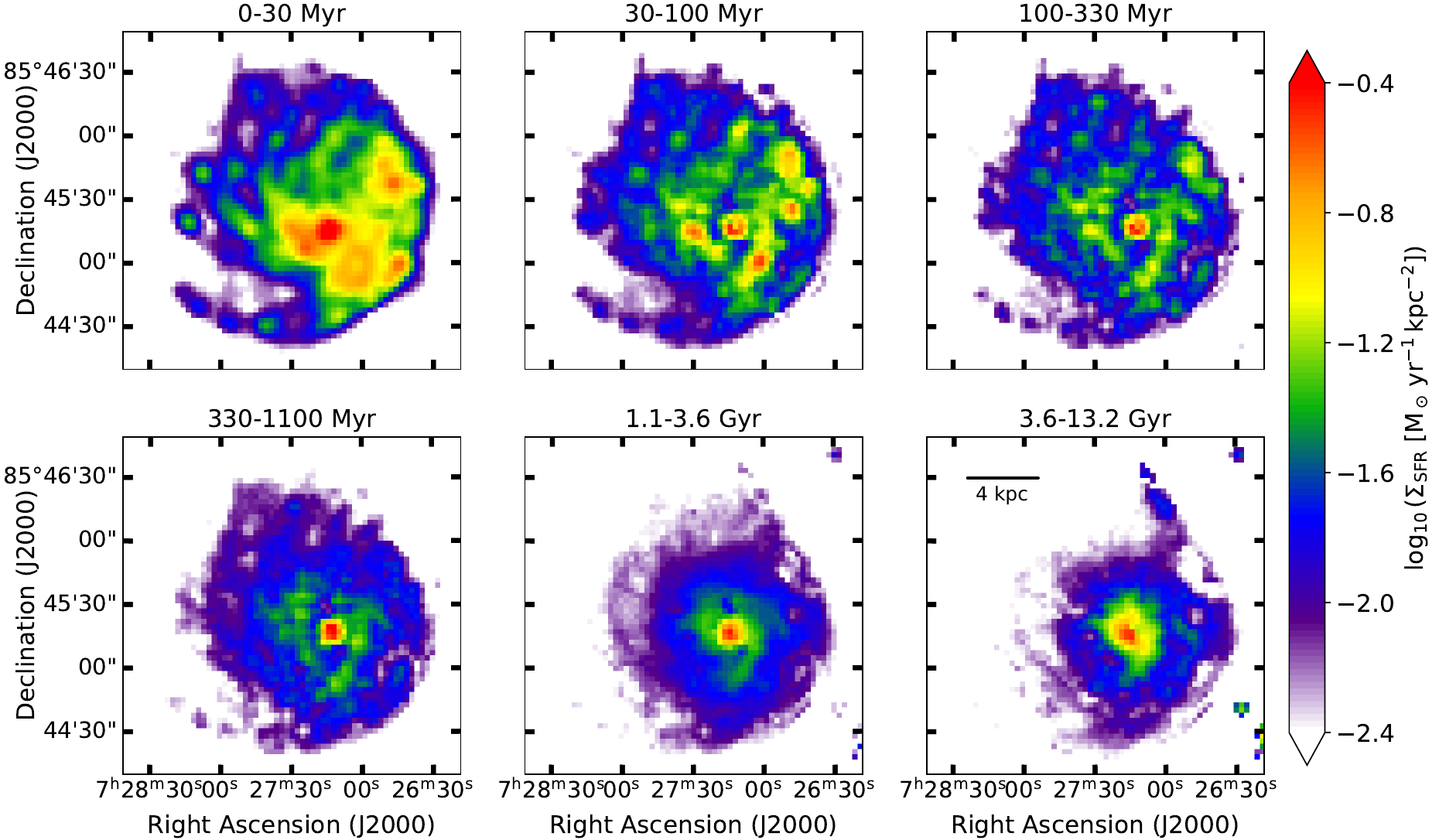}
    \caption{Star formation rate surface density as a function of position in the galaxy for different epochs corresponding to the six main age bins, derived as a \textsc{bagpipes} posterior. Details of the unit displayed in the colorbar are written in Sect. \ref{ssect:pix-by-pix}.
    }
    \label{fig:sfh}
\end{figure*} 

For all SED fitting cases, we adopted the dust model by \cite{CF00} with the fixed value of the multiplicative factor on A$_\mathrm{v}$ for stars in birth clouds, $\eta = 3$, and for the power-law slope of the attenuation law, $n$, we adopt a value of 0.7.
We make use of non-parametric SFHs, adopting 6 age bins (see Table \ref{tab:bp_priors}) that are similar to those used by \citet{Marasco+2025L} and \citet{Leja19} with the exception that we merged their two oldest age bins. Following \citet{Leja19}, we make use of a "continuity" prior that ensures smooth SFHs, preventing against sharp changes in the SFR between consecutive age bins.
All prior parameters for the spatially resolved case are listed in Table \ref{tab:bp_priors}.
We adopted the "continuity" SFH prior because of its proven flexibility in modeling highly star-forming galaxies \citep{Leja19} and its widespread adoption as a standard in SED fitting \citep[e.g.,][]{Kaushal+24, Marasco+2025L, watson+25}. Furthermore, our specific age bins were chosen because the distinguishability of simple stellar populations scales roughly with their logarithmic time separation \citep{Ocvirk+2006}. However, we had several SED fitting tests being done with different choice of SFH priors (e.g., "double power law", \citealt{Iyer+19} model, \citealt{Tacchella+22} model, \citealt{Wild2020} model) on both the integrated fluxes and by choosing several spaxels across the galactic disc. All of the models were able to either model the SFH \citep[and the most recent SFR estimates from the data, e.g.,][]{tomicic18} compatible to the one modeled by the "continuity" prior, or they resulted in a bad fit to the observed data.
Additionally, due to adding redshift as a prior to \textsc{bagpipes}, our initial SFR posteriors were calculated at the distance of $\sim35.5$ Mpc. 
By taking into account the group distance of 28.5 Mpc from \citet{fadda+23} as a fiducial distance, we had to correct our results by dividing them with the squared ratio of the two distances, as was already discussed in Sect. \ref{sect:intro}.

\subsection{Spatially resolved SFH}
\label{ssect:pix-by-pix}

In Fig. \ref{fig:fit_ex} we show an example of a fitted spectrum for the parameters derived from our SED modeling for an individual spaxel of NGC 2276.
All of the photometric points for this spatial bin are within the 16th-84th percentile of the fitted spectrum.
Three more examples of fitted spectra for pixels around the galactic disk are shown in the Appendix, as well as the corner plot for the fit from Fig. \ref{fig:fit_ex}.

In this work we focus on the SFH derived with \textsc{bagpipes}. Other fitted parameters are shown in the Appendix. Fig. \ref{fig:sfh} represents \textsc{bagpipes} posterior in the form of the star formation rate density ($\Sigma_\mathrm{SFR}$) maps\footnote{These maps are derived by dividing the \textsc{bagpipes} SFR posterior for each age bin with the area of the spatial sampling ($\sim0.4$ kpc)} for the six main age bins.
Although we call these for simplicity "SFR density map", for the oldest age bins they do not represent how the "instantaneous" star formation was spatially distributed in the past, as stars have long left their original birth location.

The figure indicates that stars of age older than 1.1 Gyrs in this galaxy are distributed fairly symmetrically. The main deviation from the symmetry appears in the oldest age bin in the northwest part, and is likely an effect of the WFC3 detector anomaly in filters F814W and F555W \citep{fowler+17, Huynh+24}, which was already described in M26. 

\begin{figure}[t!]
    \centering
    \includegraphics[width = 0.99\hsize]{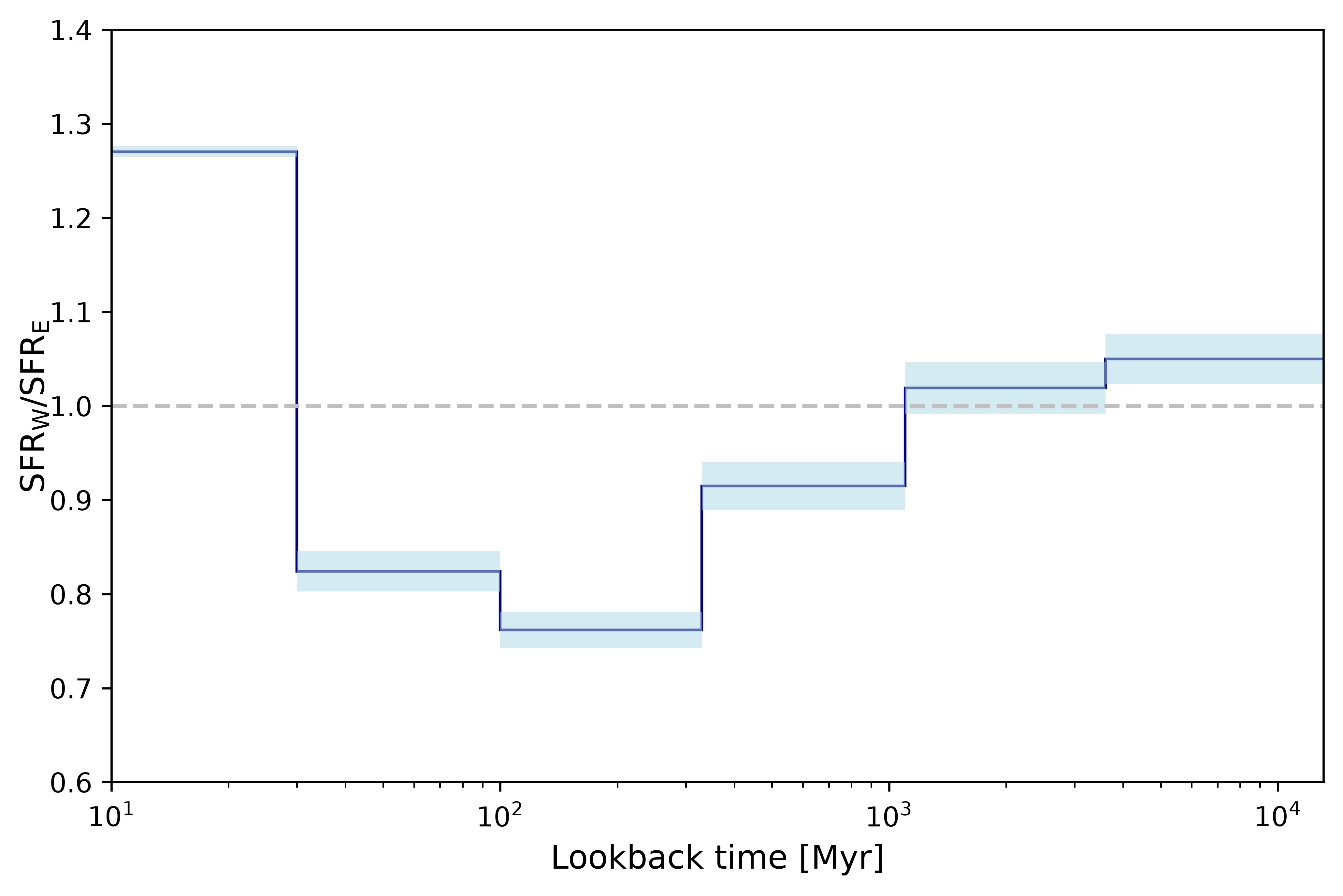}
    \caption{ Ratio of the total star formation rate on the western (leading) and the eastern (trailing) side of the galaxy across the different time epochs calculated from the spatially resolved star formation history maps. East-west asymmetry is visible solely in the young stellar populations. }
    \label{fig:sfr_ratio}
\end{figure} 

Fig. \ref{fig:sfh} clearly shows that the only stellar population responsible for the asymmetric morphology of NGC 2276 are the stars younger than $\sim1.1$ Gyr.
These stars produce both the elongation visible towards the eastern side and the high density region visible towards the western side.
By dividing the galaxy into the western (leading) and eastern (trailing) side using our division into eight sectors shown on Fig. \ref{fig:binning}, we investigated how the total SFR changes on two opposite sides of the galaxy through cosmic time.
Our results are shown on Fig. \ref{fig:sfr_ratio} and they again clearly show that only the younger stellar populations are asymmetrically distributed.
However, this plot lacks the information on the spatial variations in the SFR on smaller scales so it should interpreted carefully.
The youngest age bin indicates that SFR increases more on the western side, compared to the eastern side of the galaxy. Furthermore, areas with the highest $\Sigma_\mathrm{SFR}$ coincide with the areas of shock-enhanced $\left[\ion{C}{II}\right]$ emission \citep{fadda+23}, as well as with areas with the ultra-luminous X-ray sources (ULX) \citep{wolter+15, Anastasopoulou+19}. 
In Table \ref{tab:SFR_comp_young_stars} we compare the values of the total SFR in two of the youngest age bins for the central region (r$\lesssim12''\sim1.65$ kpc), the western, and the eastern side of the galaxy without the central region. The western and eastern regions were defined by using the eight angular sectors as shown in Fig. \ref{fig:binning}.
The results show that, while all of the selected regions experienced an increase in the total SFRs in the last $\sim30$ Myrs, the eastern side remained mostly stable ($\sim0.07$ dex increase). 
The western side shows an increase of $\sim0.2$ dex, and in the central region this increases up to $\sim0.5$ dex.

\begin{table}[h!]
\caption{Comparison of the total SFR (in units of $\mathrm{M_\odot \, yr^{-1}}$) for two of the youngest age bins for the central region (r$\leq$1.65 kpc), and the western and eastern side outside the central region.}
\label{tab:SFR_comp_young_stars}
\centering
\begin{tabular}{lll}
\hline
\hline
Region                         & SFR (0-30 Myr)                                     & SFR (30-100 Myr)          \\
\hline
Center                         & $1.67\pm0.02$                                      & $0.58_{-0.04}^{+0.05}$    \\
Western side                   & $5.71\pm0.01$                                      & $3.26_{-0.05}^{+0.07}$    \\
Eastern side                   & $3.42\pm0.02$                                      & $2.91_{-0.04}^{+0.05}$    \\
\hline
\end{tabular}
\end{table}

\subsection{Radial distribution of SFR in time}

\begin{figure}[t!] 
    \begin{tikzpicture}
        \node[anchor=south west, inner sep=0pt] (base_img) at (0,0) {
            \includegraphics[width = \hsize]{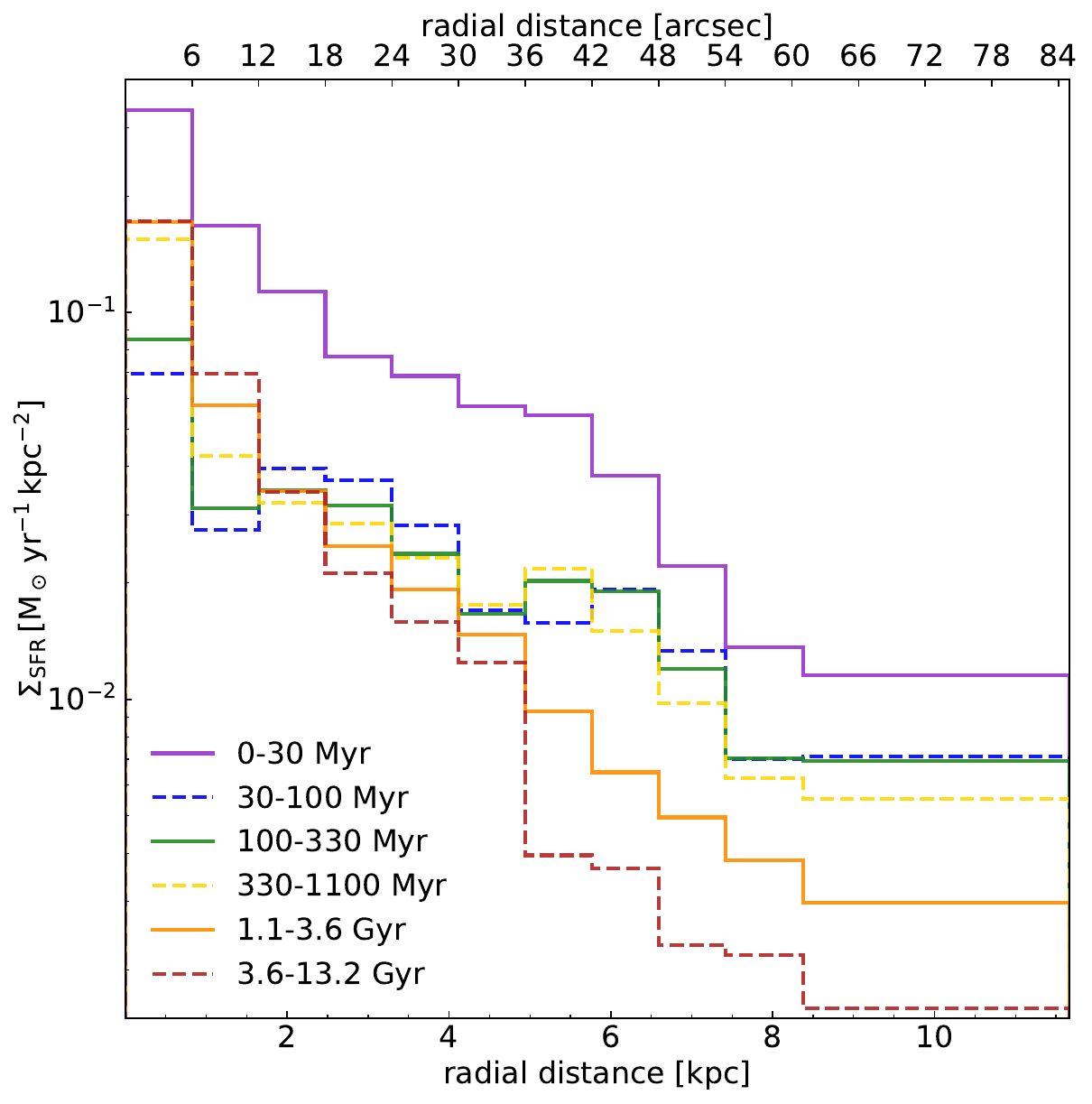}
        };

        \node[anchor=south west, inner sep=0pt] at (3.8cm,5.8cm) {
            \includegraphics[width=5cm]{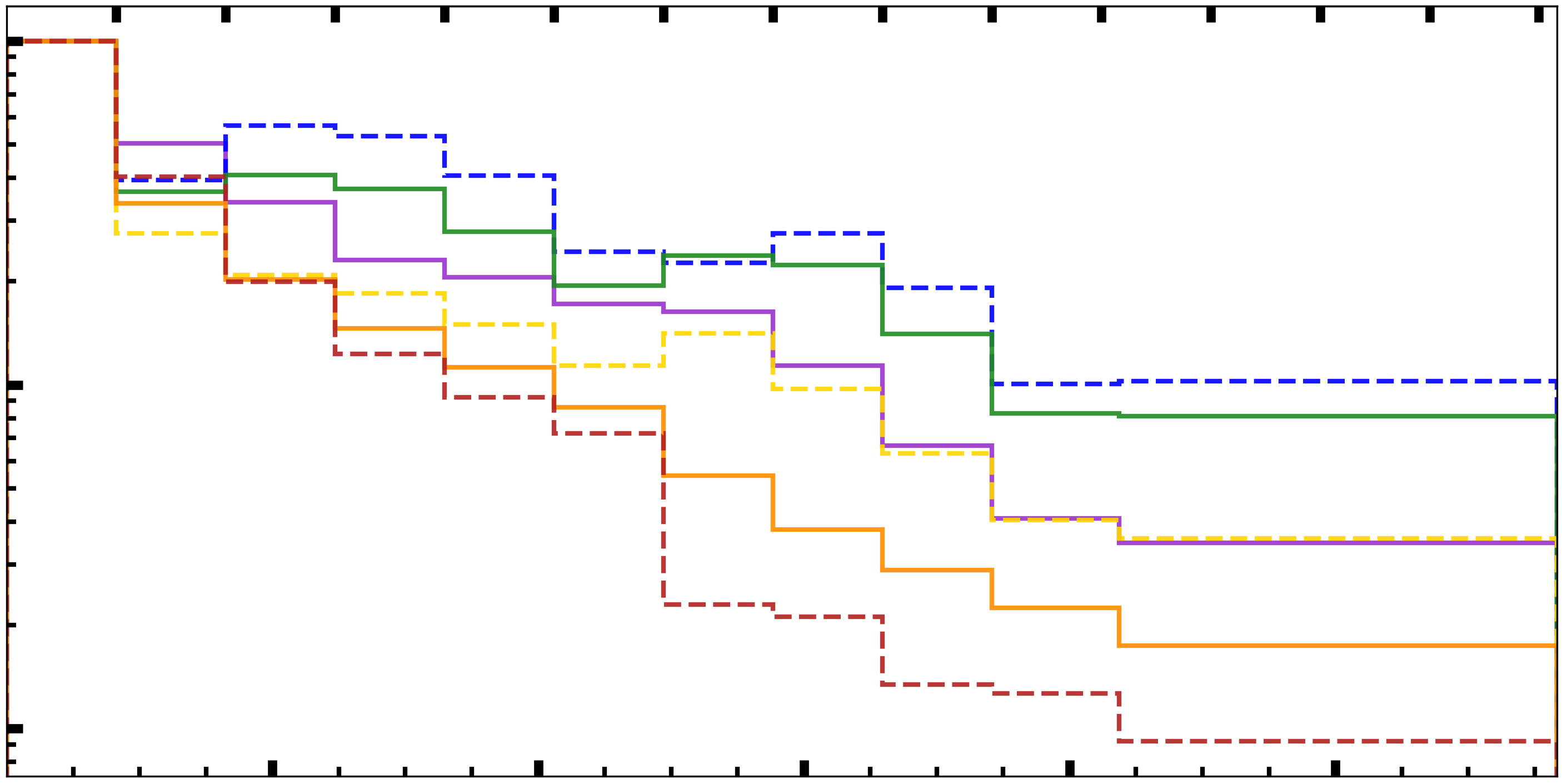}
        };

    \end{tikzpicture}
    \caption{Star formation rate surface density as a function of radial distance from the galactic center for different epochs corresponding to the six main age bins, derived as a \textsc{bagpipes} posterior. Upper right panel shows the same distribution normalized to the value of the innermost bin. }
    \label{fig:sfh_rad}
\end{figure}

As mentioned above, along with the fully spatially resolved SED fitting, we have also done SED fitting on our flux maps which were binned into concentric ellipses \citep[assuming an inclination of $\sim20$°;][]{tomicic18} with a minimum radial separation of 6$''$. We show the results on the temporal evolution of the $\Sigma_\mathrm{SFR}$ radial profile in Fig \ref{fig:sfh_rad}.
The goal of this approach was to see if there are differences in the radial distribution of the SFR for different age bins. 
This approach is meant to derive more robust estimates for the SFR radial profile at different epochs, under the simplifying assumption of pure circular orbits (that is, no radial migration for the stars). 

The resulting radial distribution of the $\Sigma_\mathrm{SFR}$ shows a steady increase towards the center throughout cosmic time, with the general radial trend slightly flattening towards younger age bins.
When comparing the SFR in the last 30 Myrs, there is a visible increase at all distances from the center compared to the older age bins, as can also be seen from the spatially resolved SFH shown in Fig. \ref{fig:sfh} and from Table \ref{tab:SFR_comp_young_stars}. 
Additionally, it is visible that the SFR increases in the outer parts of the galaxy when going from the older towards younger ages, which shows the increase in the size of the galactic disk.

\subsection{Total SFH}

We now present the total SFH of NGC 2276 computed using different configurations, and compare these with the SFH of our NGC 2276-like hydrodynamical simulation, described in details in Sect. \ref{sect:sim}). 

\begin{figure}[t!]
    \centering
    \includegraphics[width = 0.99\hsize]{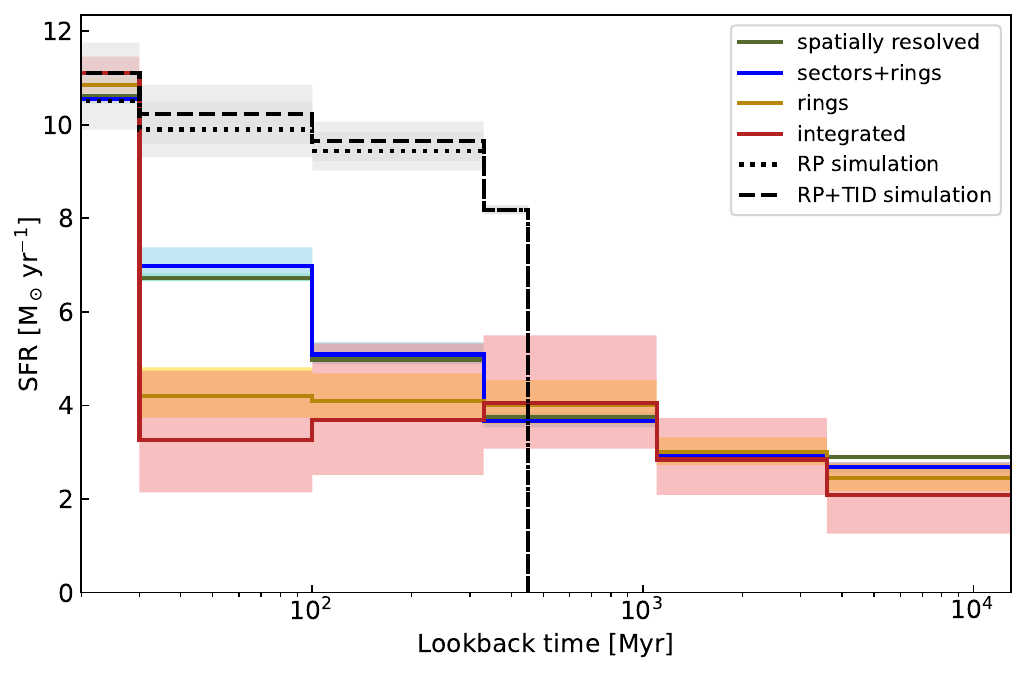}
    \caption{Star formation rate as a function of lookback time for all differently resolved cases, compared with the simulated SFH. 
    }
    \label{fig:sfh_comp}
\end{figure}

In Fig. \ref{fig:sfh_comp}, we show the total SFH of NGC 2276 determined with the various configurations: integrated SED modeling (red), individual elliptical radial bins (gold), individual radial bins further divided into 8 sectors (blue) and individual pixels (green). 
From the different \textsc{bagpipes} fits, we obtained the value of ongoing SFR (in the last 30 Myrs) to be $\sim10.8_{-0.5}^{+0.4}$ M$_\odot/$yr, which is comparable to the value derived by \cite{tomicic18} from the integrated and attenuation corrected H$\alpha$ emission \citep[$\sim12\pm4$ M$_\odot/$yr, corrected for distance of the NGC 2300 group from][]{fadda+23}. 

The main feature that emerges from the total SFH is the factor $\sim1.6$ jump in SFR between the two most recent age bins (0-30 and 30-100 Myr) in the most resolved configurations. 
The difference between these two measurements is expected to arise from the different timescales traced by the H$\alpha$ ($\sim$few tens of Myr) and the UV+MIR bands ($\sim$100 Myr) and could indicate a very recent episode of increased star formation. 
Additionally, this $\sim1.6$ factor was also visible in the work of \citet{tomicic18}, where the authors also calculated the SFR from FUV+22$\mu$m and derived a value of $\sim7$ M$_\odot/$yr \citep[corrected for distance of the NGC 2300 group from ][]{fadda+23}, which matches very well with our derived SFH. 
This factor between the two most recent age bins increases to $\sim2.8$, for the least resolved configurations. 
This discrepancy could be due to the more integrated cases being less sensitive to changes in the spatial variations of the dust on smaller scales, or due to more integrated configurations underestimating total galaxy mass \citep[e.g.,][]{sorba+15} where in our example there was a factor of $\sim1.6$ difference in the total stellar mass between the most and the least resolved configurations. 
This shows how differently resolved data can impact the results of SED modeling and the interpretation of the results of SFH, and that using more resolved data can be beneficial in studying the evolution of galaxies.
Additionally, we checked if our results for SFH for the nearest age bins (up to 100 Myrs) differ much when we avoid using the emission line maps. However, our results stayed roughly the same, but the extent of the errors for the nearest age bin has largely increased, as expected.

\section{Simulation}
\label{sect:sim}

Here we make use of idealised hydrodynamical simulations to better assess the impact of ram pressure alone, and of a combination of ram pressure and gravitational interactions, on the evolution of a NGC 2276-like system. 
Our models are not meant to be an exact replica of the NGC2300 group environment, but rather controlled numerical experiments made to infer the main differences in the morphology and kinematics of the stellar and gas components caused by the different environmental mechanisms

\subsection{Galaxy setup}

The initial conditions of our model disk galaxies are built using the publicly available initial conditions set-up code Disk Initial Conditions Environment \citep[DICE,][]{perret+14}.
DICE builds a composite model galaxy including a dark matter halo, stellar disk, and gas disk component.
By measuring the shape of the gravitational potential from the total model computed on a multi-level grid, it assigns the positions and velocities of dark matter and star particles to try to produce an overall model where the different components are in equilibrium with each other. 
We find that this produces initial conditions that are already very stable. 
To test for stability, we monitored the evolution of the 20\% and 95\% Lagrangian radii, which trace the inner and outer regions of the stellar disk and dark matter halo, respectively. 
Models were considered stable if these radii remained constant over several crossing times of the system.

We model NGC 2276 as an exponential stellar disk with M$_*=4\times10^{10}$ M$_\odot$, and an effective radius of 4 kpc.
The model consists of $10^5$ star particles initially within the plane of the disk in the x-y plane, yielding a particle mass of $4\times10^5$ M$_\odot$. 
To model the dark matter halo, we use the Navarro, Frenk and White (NFW) profile \citep{NFW+97} consisting of $10^6$ dark matter particles, with particle mass of $1.2\times10^6$ M$_\odot$. 

For tidal interaction models we include the secondary galaxy NGC 2300. We model its profile as an early-type de Vaucouleurs stellar component modeled with a Hernquist sphere \citep{hernquist+1990} of M$_*=9.6\times10^{10}$ M$_\odot$, with effective radius of 4 kpc.
We modeled the stellar component with $10^5$ star particles, and embedded it in an NFW halo of $10^6$ dark matter particles. 

From abundance matching \citep{guo+2010} we extract the halo mass of $1.2\times10^{12}$ M$_\odot$ for NGC 2276 model, and $5.8\times10^{12}$ M$_\odot$ for NGC 2300 model.
From the halo-concentration relation \citep{child+17} we assume a concentration parameter c=7 for NGC 2276, and c=6 for NGC 2300 from their respective halo masses. 
Pre-RPS gas mass for NGC 2276 is $7\times 10^9$ M$_\odot$ \citep{cortese+21}, and we assume that the gas disk is more extended than the stellar component with an exponential scalelength that is 1.7 times larger \citep{cayatte+94}. 

\subsection{Interaction parameters}

Fiducial interaction is set up with a separation of 250 kpc in y-dimension, and separation of 50 kpc in x-dimension. 
The spiral is placed stationary, and the elliptical approaches it with a negative y-velocity of 900 km/s (see below for a justification of these values). 
Thus the disk is approached from the edge-on direction but the initial x-separation means the collision is not direct, and the elliptical passes the disk in a retrograde direction. 
This choice places the elliptical below the ram pressure stripped gas tail, as observed. 

By slightly rotating our line of sight (by 20-30 degrees about the y-axis), we find we can provide a reasonable match to the observed relative velocity of the galaxies in line of sight \citep[$\sim500$ km/s,][]{fadda+23}, separation on the plane of the sky ($\sim52$ kpc), and relative location on the sky down our line-of-sight after NGC 2300 has passed NGC 2276.
Observations confirm that the disk is slightly inclined to our line of sight \citep{tomicic18}. 

We additionally tested models with an x-separation of 10 kpc and 25 kpc, which results in a closer passage. 
However, we found that the resulting tidal interaction resulted in a highly disturbed stellar disk morphology (with strong barring, induced grand-design spirals, and some overall asymmetry) which was not a good match to the observed morphology of NGC 2276, and yet did not significantly change the morphology of the stripped gas stream.
That is why we chose the model with the 50 kpc separation as the fiducial RPS+tidal configuration.

\subsection{Simulation code} 

Our initial conditions are evolved with the adaptive mesh refinement code Ramses \citep{teyssier+2002}. 
We chose a box size of 400 kpc which is sufficient to contain the model disk galaxy and elliptical companion. 
The boundary conditions are outflow-like, to allow the flow of intragroup medium in the wind tunnel runs to smoothly exit the box. 
The minimum refinement level is 6, and the maximum refinement level is 12, meaning a minimum cell size is 100 pc.

Adaptive mesh refinement is controlled by multiple criteria, including the number of dark matter and star particles from the initial conditions that are found in a cell (we choose 50 for all levels), and the baryonic (stars and gas) mass in a cell which depends on the refinement level of the cell (at the highest level of refinement this corresponds to a mass of 700 M$_\odot$). 

Additionally, refinement occurs to ensure the Jeans length is resolved to a factor of 4, up until the maximum refinement level. 
This choice of refinement criteria ensures that the gas and stellar disk is resolved to the maximum level of refinement
out to the edge of the disk and also vertically out of the plane of the disk. 

Simulations are conducted for 450 Myr with a snapshot interval of 14.9 Myr. 
This is sufficient time for the secondary galaxy to complete a fly by and perturb the disk in tidal interaction runs, and for a gas tail to form and gas to be stripped down to the truncation radius in ram pressure stripping runs.

Our star formation and feedback recipe is based on recipes implemented in \citet{perret+14}. 
Star formation occurs for gas above a density threshold of n$_\mathrm{H}= 0.1$ cm$^{-3}$ \citep{springel+2003}, and follows a Schmidt relation \citep{kennicutt+98} dependency on gas density with a star formation efficiency of 0.03. 
This choice of parameters ensures our isolated model galaxy forms stars at a rate that is consistent with the star forming main sequence at z=0 as in \citet{speagle+2014}. 

Supernova feedback from OB stars is treated as follows. 
The OB type stellar populations that reach an age of 10 Myr are transformed into supernovae, and release energy, mass and metals into the nearest gas cell. 
The energy release is 10$^{51}$ erg per 10 M$_\odot$ supernovae, and the release is treated thermally. 
20\% of the mass of the star particle that goes supernova is returned to the gas with a yield of 0.1.
Gas cooling occurs radiatively \citep{sutherland+93}, assuming a gas metallicity of 0.15 solar metallicities down to a temperature floor. 
The temperature floor is dictated by the Truelove condition \citep{truelove+97} to guard against artificial fragmentation of the gas. 
We use the default background ultraviolet heating model of Ramses for redshift=0 \citep{haardt+96}.

\subsection{Wind tunnel simulations}

In order to model the impact of ram pressure stripping on our model galaxy, we use a so-called ‘wind-tunnel’ approach.
The model galaxy is static within the simulation volume and a hot ionised gas wind is fed into the simulation volume from a specified region, with a chosen density and wind speed, mimicking the motion of the disk galaxy through an ambient gas, such as the intracluster medium of a cluster. 
In this study, the region from which the wind is fed in is located at 125-150 kpc in the x-direction, and covers almost the full y-z face of cube (10-390 kpc in the y and z direction). 
The wind is fed in along the x-direction, with a fixed gas density of $5\times10^{-27}$ g/cm$^{3}$ \citep{davis+96, rasmussen+2006}, a fixed velocity of v$_x$=-900 km/s \citep{roberts+24} resulting in a Mach number of $\sim2$, and a temperature of $10^7$ K (i.e., a constant wind model). 
As the disk of NGC 2276 lies in the x-y plane, this results in an edge-on disk stripping.

\subsection{Analysis of simulated data}

\begin{figure}[h!]
    \centering
    \includegraphics[width = 0.99\hsize]{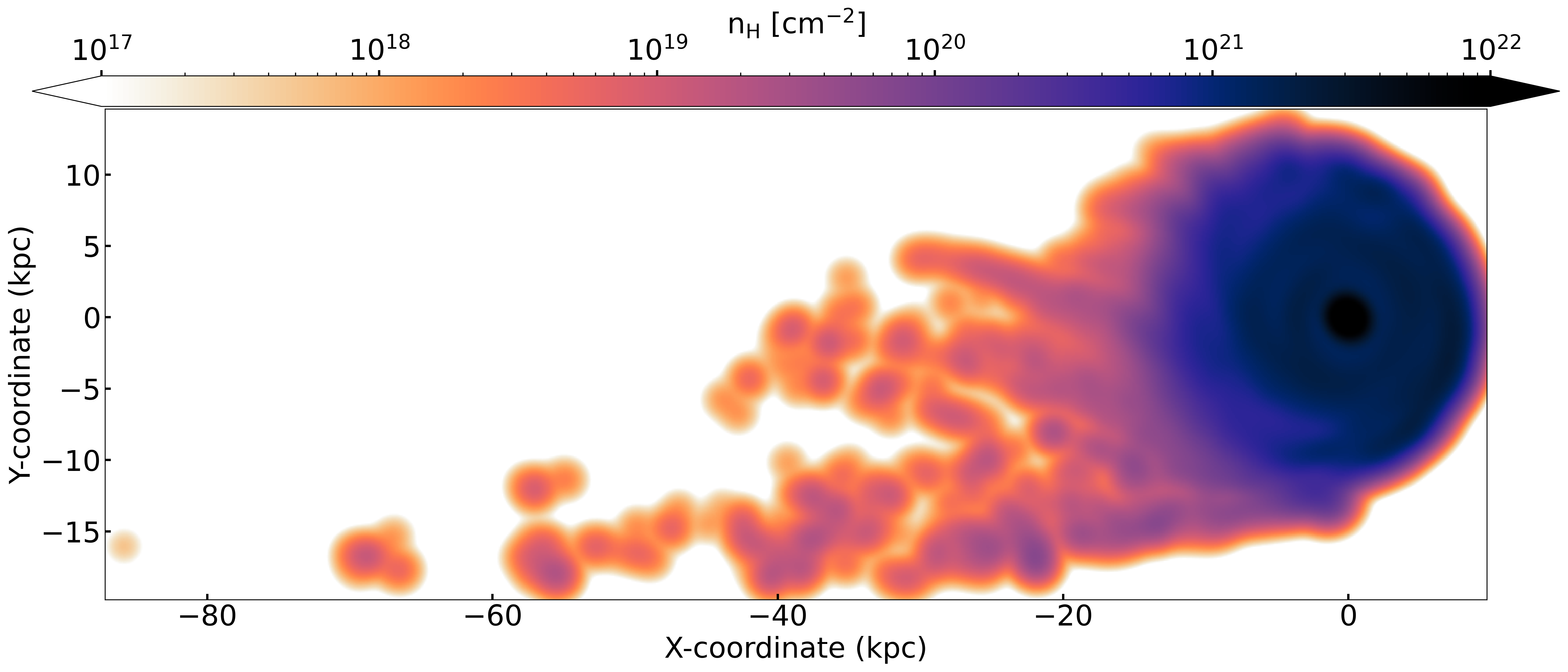}
    \includegraphics[width = 0.99\hsize]{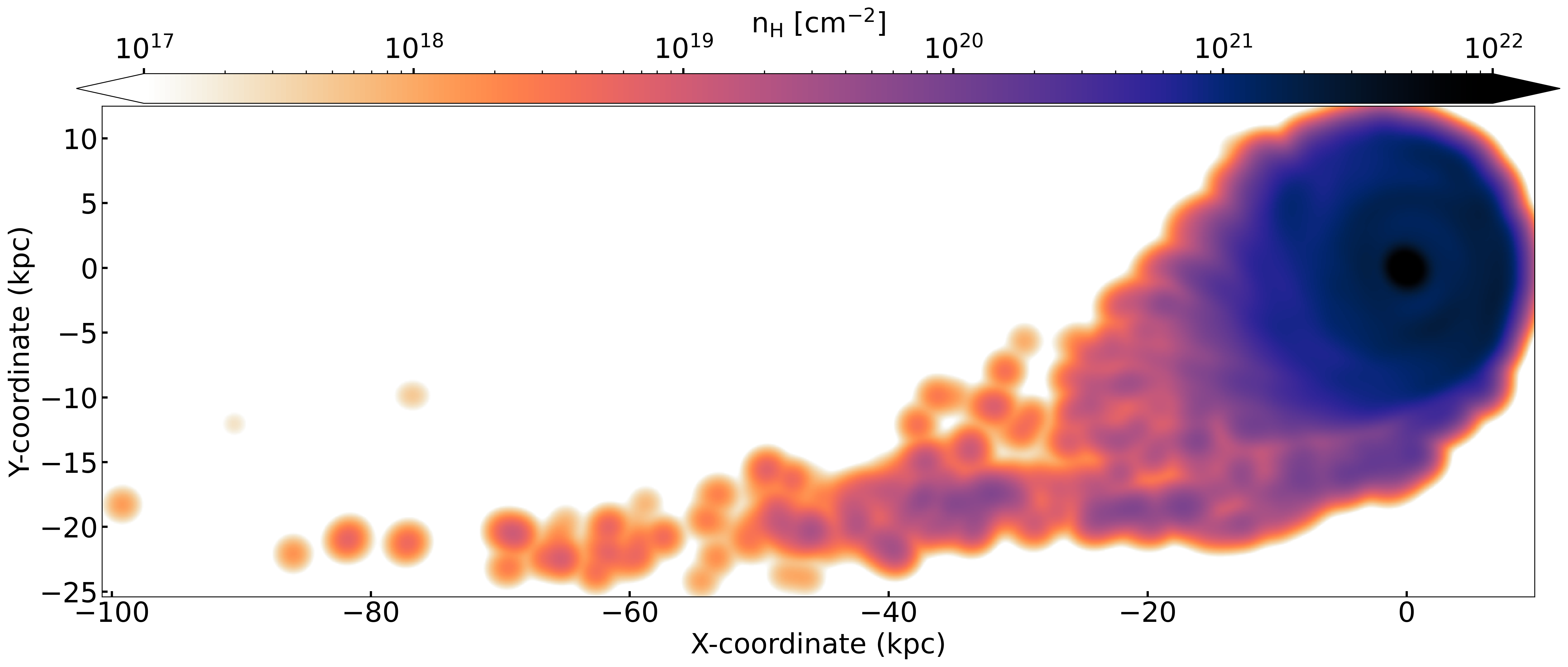}
    \includegraphics[width = 0.99\hsize]{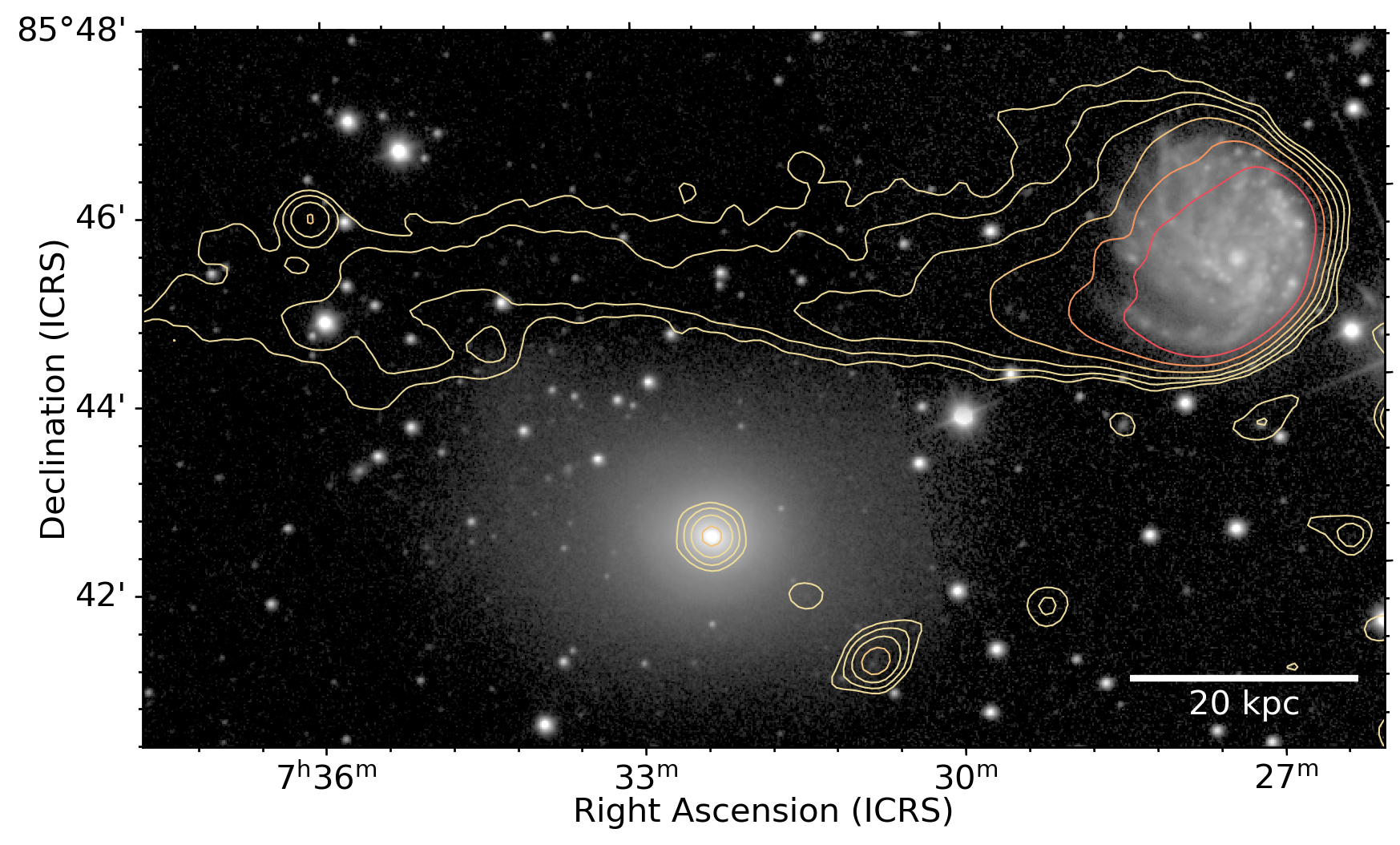}
    \caption{Top: Gas column density map for the RPS+tidal simulated case. Middle: Gas column density map for the RPS simulated case. Only gas with density n$_\mathrm{H}>0.1$ cm$^{-3}$ is shown, and the spatial resolution of the map is adjusted to match that of the LOFAR HBA low resolution image of NGC 2276 \citep{roberts+24}. The gas columns density maps are in x-y plane and are not projected to match the viewpoint that matches the inclination of NGC 2276.
    Bottom: LOFAR HBA 144 MHz emission contours (6, 12, 24, 48, 96, 192$\times$RMS, RMS = 120$\mu$Jy, resolution = 11.9 $\times$ 8.2 arcsec$^2$ ) plotted on top of Palomar's Zwicky Transient Facility g-band image of NGC 2300 group center.
    }
    \label{fig:gas_sims}
\end{figure}

In the top and middle panels of Fig. \ref{fig:gas_sims} we show the simulated gas component of a NGC 2276-like system from our simulation of RPS with tidal interaction (with fiducial 50 kpc initial separation), and RPS-only experiments, respectively. 
In both configurations, there is visible gas compression on the leading side, as well as clear gas stripping on the trailing side.
The simulated tail length is comparable to what was observed from synchrotron emission in \citet{roberts+24}, as shown in the bottom panel.
We stress that this comparison is purely qualitative, given that our simulation does not include the physics required to predict non-thermal radio emission (e.g., magnetic fields and relativistic electron transport). 
Yet, regardless of the physical mechanism responsible for the emission, there are remarkable similarities between the dense gas distribution in the simulation and the observed radio-emitting tail, supported by the fact that the thermal and nonthermal stripped ISM typically show a close resemblance \citep{Ignesti2022, foster+26}.
We notice that the upper-left extension of the gaseous component (northeastern part of the observed galaxy) is more prominent in the RPS+tidal configuration, than in the RPS only. 
That effect could potentially be due to the specific configuration of our RPS+tidal model. 
As the NGC 2300 model passes NGC 2276 model, it tidally pulls it downwards by a certain amount, and the result is the RPS is no longer purely operating from right to left, but has a slight vertical component as well, which seems to slightly improve the tail morphology.
This feature is potentially similar to the observed northeastern feature observed in the synchrotron tail, shown on Fig. \ref{fig:gas_sims}.

\begin{figure}[t!]
    \centering
    \includegraphics[width = 0.99\hsize]{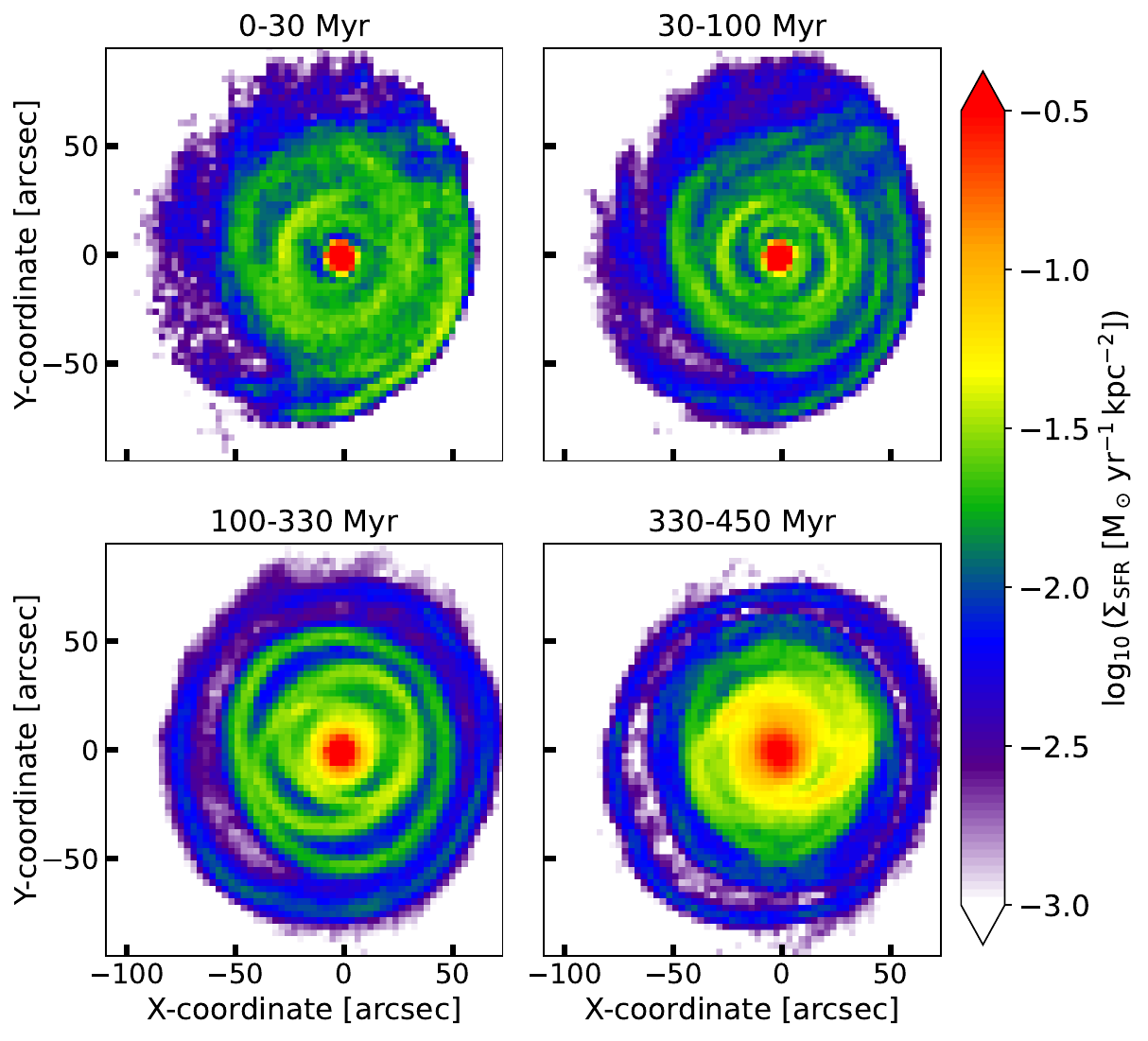}
    \caption{Star formation rate surface density maps for the simulated galaxy with RPS+tidal interaction in four different age bins. The galaxy is seen from a viewpoint that matches the inclination of NGC 2276.
    }
    \label{fig:sfh_sims_rp+tid}
\end{figure} 

\begin{figure}[t!]
    \centering
    \includegraphics[width = 0.99\hsize]{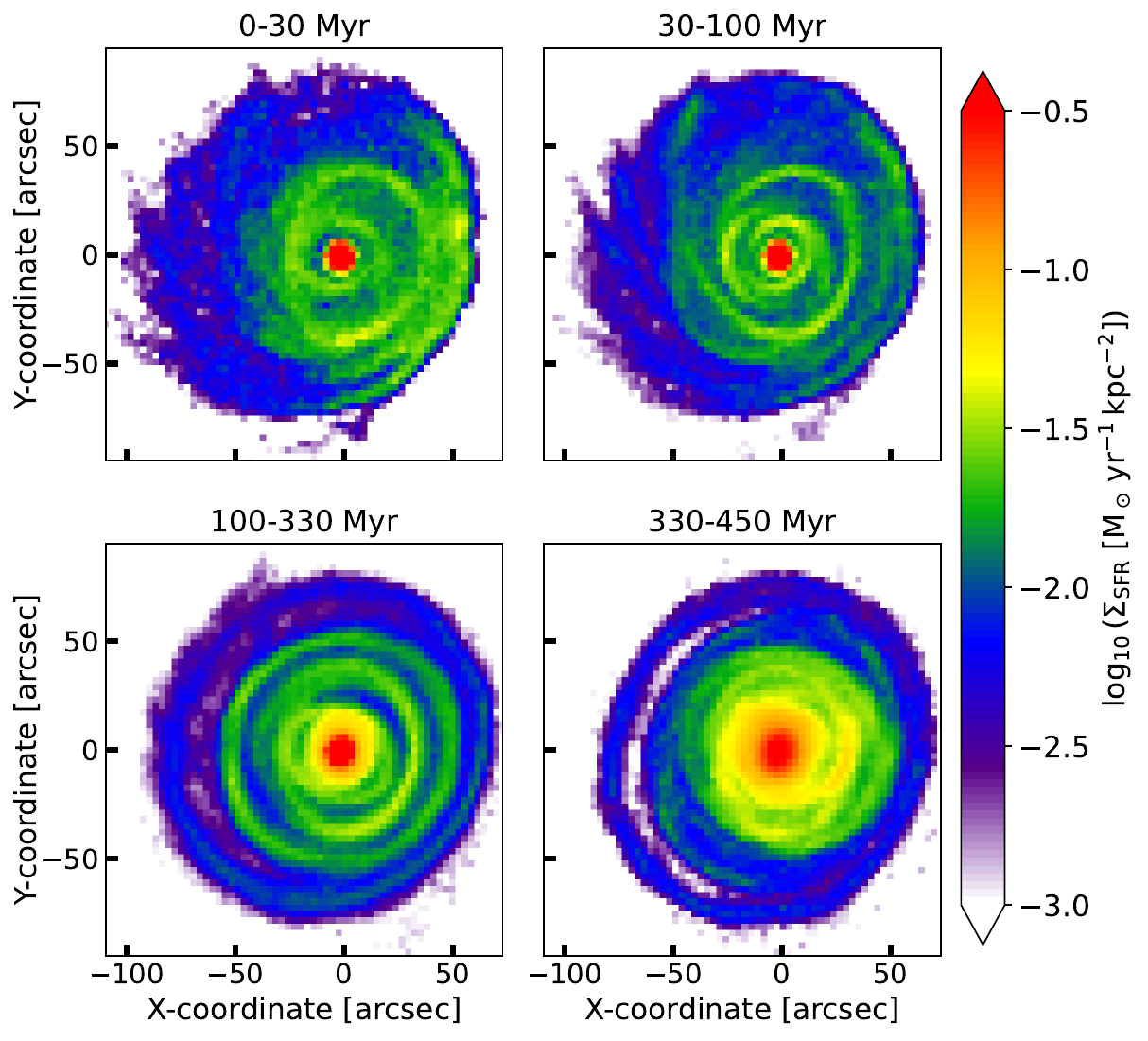}
    \caption{Star formation rate surface density maps for the RPS-only simulated galaxy in four different age bins. The galaxy is seen from a viewpoint that matches the inclination of NGC 2276.
    }
    \label{fig:sfh_sims_rp}
\end{figure} 

In Fig. \ref{fig:sfh_sims_rp+tid} and Fig. \ref{fig:sfh_sims_rp} we show maps of $\Sigma_\mathrm{SFR}$ calculated by dividing the total stellar mass formed in the simulation in particular age bin (per pixel area of $0.4\times0.4$ kpc$^2$) with the width of each of the age bins. 
The maps are projected to match the inclination of NGC 2276 of $\sim20$° \citep{tomicic18}. 
The spatial distribution is mostly symmetrical in both the RPS-only and the RPS+tidal configuration until the most recent 100 Myrs, when the first clear sign of unwinding appears. 
However, the $t < 30 \, \mathrm{Myr}$ age bin in both configurations shows that the star formation is mostly concentrated within the leading galaxy side, compared to the trailing side.
We stress that these results do not show the realistic simulated SFH of NGC 2276-like system in terms of absolute values of SFR, but they should rather be examined through differences in SFR between adjacent time bins since the predicted times of the impact with the IGM.

In Table \ref{tab:SFR_sim_comp_young_stars} we show comparison of the median $\Sigma_\mathrm{SFR}$ for both configurations of the simulated galaxy on eastern and western sides of the galactic disk. 
For determining the eastern and western sides we use the same sectors as for calculations used in Table \ref{tab:SFR_comp_young_stars}.
Both configurations show similar median values of $\Sigma_\mathrm{SFR}$. 
Although the general distribution of SFR across all 4 age bins looks similar, with a distinct east-west asymmetry in the nearest age bin, the RPS-only simulation shows a much extended eastern side in the younger age bins compared to the RPS+tidal configuration.

\begin{table}[h!]
\caption{Comparison of the median $\Sigma_\mathrm{SFR}$ (in units of $\mathrm{M_\odot \, yr^{-1} \, kpc^{-2}}$) for the youngest age bin ($t < 30 \, \mathrm{Myr}$) for the western and eastern side outside of the two simulated galaxy configurations.}
\label{tab:SFR_sim_comp_young_stars}
\centering
\begin{tabular}{lll}
\hline
\hline
Region                         & $\overline{\Sigma_\mathrm{SFR}}$ (RPS+tidal)        & $\overline{\Sigma_\mathrm{SFR}}$ (RP only)          \\
\hline
Western side                   & $0.016\pm0.006$                                      & $0.014\pm0.005$                                    \\
Eastern side                   & $0.004\pm0.001$                                      & $0.006\pm0.001$                                    \\
\hline
\end{tabular}
\end{table}

From Fig. \ref{fig:sfh_comp} it is visible that the overall shape of the SFH from our simulations looks roughly the same. The total values are greater when RPS and tidal interaction are working in cooperation. However, it is interesting to observe that, even though the SFHs from the simulations show a gradual growth towards younger ages, they do not fully replicate the total SFH from either of the four configurations of SFH derived with \textsc{bagpipes}, but it manages to replicate the current SFR measured within the latest 30 Myrs.

\section{Discussion}
\label{sect:discussion}

In our previous work M26, we showed that the morphology of NGC 2276 was most likely disturbed by interacting with its IGM via ram pressure. 
We presented evidence of ionized gas (H$\alpha$) and young stellar disk truncation on the leading side, as well as the extended morphology of the same tracers on the trailing side. 
Additionally, we also demonstrated that the older stellar disk is extended $\sim2$ kpc further than the young stellar disk on the leading side. 
However, due to our analysis being performed only on the broadband filters from HST and Spitzer, we could not make any straightforward conclusions on whether ram pressure is the only disturber of this system due to the relatively close projected distance ($\sim52$ kpc) to its companion NGC 2300.

In this work we have been able to make a better distinction between the stellar populations by using the spatially resolved SED fitting. 
In Fig. \ref{fig:sfh} we show that the older stellar populations ($t > 1.1 \, \mathrm{Gyr}$) are mostly symmetrical, while the younger ones show an increased amount of asymmetry, as well as extended morphology on the easternmost side. 
This is especially pronounced in the youngest stellar population ($t < 30 \, \mathrm{Myr}$) which show a high degree of asymmetry when comparing the leading and the trailing side (as shown in Table \ref{tab:SFR_comp_young_stars}). 

Considering that the eastern side of the galaxy starts to become more extended during the $330 \, \mathrm{Myr} < t < 1.1 \, \mathrm{Gyr}$ age bin, we are inclined to believe that the RPS started during that time, which is consistent with the 1-2 Gyrs discussed by \citet{davis+96} and \citet{rasmussen+2006}. 
The similar claim cannot be made if we take into account only the SFH from Fig. \ref{fig:sfh_comp}, proving that spatially resolved analysis is the key to obtaining our results. Additionally, the same figure shows no "bursts" of SFR, prior to one in the youngest stellar populations, which could be due to the galaxy currently being near the center of the IGM, and probably at the peak of RPS \citep{roberts+24}.

Due to the fact that the older stellar populations are not morphologically disturbed, we are more confident that the only relevant large-scale effect shaping NGC 2276 is ram pressure, as ram pressure primarily affects the galaxy's gas component, including the star-forming interstellar medium (ISM), and subsequently the younger stellar populations \citep[e.g.,][]{boselli+22}. 

This was also indicated from the simulation results (shown on Fig. \ref{fig:sfh_sims_rp+tid} and Fig. \ref{fig:sfh_sims_rp}, as well as in Table \ref{tab:SFR_sim_comp_young_stars}), where there is a visible asymmetry in the youngest stellar populations, although to a somewhat smaller degree. 
Both simulation configurations (RPS-only and RPS+tidal) show similar spatial distribution of both younger and older stellar populations, which indicates that tidal interaction had a minor effect on the morphology of this system in this simulation setup, if any.
The regions of higher $\Sigma_\mathrm{SFR}$ in the youngest age bin for the simulated galaxy (top-left panel in Fig. \ref{fig:sfh_sims_rp}) and for NGC 2276 (top-left panel of Fig. \ref{fig:sfh}) appear to match remarkably well. 

It is important to mention that we have only made these comparisons between the two simulation configurations, RPS-only and RPS+tidal with an initial separation of the two galaxy models of 50 kpc.
However, we have tried to simulate RPS+tidal cases with different initial separations, 25 kpc and 10 kpc.
In both of those cases we start to see some of the features that are not present in the observed data of NGC 2276 like strong barring and the distortion of the stellar disk, which is why we ruled out those configurations and decided to continue our analysis with the model that has the initial separation of 50 kpc as the fiducial one.

Additionally, from Fig. \ref{fig:sfh_comp} it is visible that there is a discrepancy between the total SFR obtained from the simulations compared to the \textsc{bagpipes} results. 
This difference could potentially be explained with an increase in the IGM density in the recent past as NGC 2276 approaches the group X-ray emission \citet{Mulchaey+93}. 
Considering that in our simulations we use a single value for the galaxy velocity and IGM density, it is possible that due to the recent passing of the galaxy through a more dense IGM, its SFR increases rapidly. 
This was also shown in simulations with variable edge-on ram pressure \citep{Bekki+14}, where it was shown that SFR can be enhanced up to 0.5 dex ($\sim3$ times) in $\sim100$ Myrs since the passage of the cluster center.
Another reason for this difference might arise from the possible age-metallicity degeneracy in our SED fitting, which can be seen from the metallicity maps shown on Fig. \ref{fig:bagpipes_ancillary} or due to some other assumption made in either the SED modeling or initial simulation conditions.

Additionally, a large jump (factor of $\sim1.6$) was found in the SFR between the $t < 30 \, \mathrm{Myr}$ and $30 \, \mathrm{Myr} < t < 100 \, \mathrm{Myr}$ age bin in the more spatially resolved configurations of the SED fits.
This is a direct consequence of the different SFR timescales traced by the H$\alpha$ ($\sim$few tens of Myr) and the UV+MIR bands ($\sim$100 Myr) as was discussed in \citet{tomicic+24}, and implies a SF burst in the recent past.
The $\sim0.3$ dex difference between the $30 \, \mathrm{Myr} < t < 100 \, \mathrm{Myr}$ age bins for the least spatially resolved and the most spatially resolved configurations could be explained due to SFR models being more sensitive to attenuation effects due to dust at smaller spatial scales \citep[e.g.,][]{tomicic+19, Belfiore+23}.
In general terms, galaxy parameters derived from modeling of the global SED might be less reliable than those derived using more resolved configurations, to the former case being weighted to the most luminous stellar populations \citep[e.g.,][]{sorba+15}.

The recent and relatively steep increase in the SFR that places NGC 2276 at 0.6 dex above the main sequence of star formation \citep[e.g.,][]{renzini+15, cano-diaz2016} might seem unusual, but not unrealistically extreme.
There are other galaxies that exhibit similar or even greater increase in the SFR. 
For example, galaxy ESO 338-IG04, although with a significantly smaller mass \citep[M$_*=1.6\times10^8$ M$_\odot$,][]{Bik+18} compared to NGC 2276, exhibits an even more drastic SFR which is $\sim$1.6 dex above the main sequence of star-formation \citep[e.g.,][]{renzini+15}. 
Similar result for the same galaxy was also shown by \citet{Ostlin+2001}, where the authors derived a different mass estimate, which resulted with an SFR of $\sim$0.8 dex above the main sequence of star-formation. 
Its starburst epoch is shown to be relatively recent \citep[started 40-50 Myrs ago and peaked $\sim10$ Myrs ago,][]{ostlin+2003}, as it also seems to be the case for NGC 2276.
Even though the origin of the starburst in ESO 338-IG04 was possibly a recent merger \citep{cannon+04}, it provides a valuable example for a galaxy featuring a SF burst with timescale and relative amplitude comparable to those found in this study. Furthermore, from our Fig. \ref{fig:sfh_comp} it is visible that in our simulations, the total SFR is higher in RPS+tidal configuration which could indicate that possible tidal interaction could trigger an additional starburst in NGC 2276-like system, increasing its most recent SFR as hypothesised by \citet{tomicic18}.

ESO 338-IG04 additionally exhibits several ULXs \citep{oskinova+19}, possibly due to its SFR that much larger for its mass. As was shown by \citet{mapelli+2010}, star-forming galaxies in the local Universe have $\sim1.3$ ULXs per unit SFR $\left(\mathrm{M_\odot/yr}\right)$, and thus unusually high SFR in NGC 2276 should not be surprising due to 11 ULXs which were found in NGC 2276 \citep{Anastasopoulou+19}, together with a significant amount of atomic and molecular gas \citep[each contributing with $\sim6\times10^9\, \mathrm{M_\odot}$,][]{rasmussen+2006, tomicic18}, as well as an increased SFE towards the western edge \citep{tomicic18}.

\section{Summary}
\label{sect:summary}

NGC 2276 is a nearby galaxy in NGC 2300 group. 
The origin of its lopsided morphology has long been debated and it was inconclusive whether it is due to ram pressure or a remnant of a tidal interaction with NGC 2300.

In this work we have carefully separated the stellar populations of different ages in this galaxy by applying the spatially resolved SED fitting to a collection of broad- and narrow-band images spanning wavelength ranges from FUV to MIR. 
From the SFH maps on sub-kpc scales we were able to conclude that stars older than 1.1 Gyrs are distributed almost perfectly symmetrical and seem to be undisturbed. 
Youngest stars are, on the other way, distributed highly asymmetrically. 
The different morphology exhibited by the old and the young stellar population is the cleanest evidence of there being negligible tidal interaction happening in the past in this system.
As for when the interaction between NGC 2276 and its surrounding IGM began, our results indicate that it happened somewhere between 330 Myrs and 1.1 Gyrs, which is consistent with previous literature sources.

Analyzing the stellar population morphology from the two different configurations of simulations (RPS-only and RPS with tidal interaction) yielded similar results, but not as clearly visible. 
Older stars were again more symmetrically distributed, and the younger ones showed asymmetry. 
There were not many differences between the RPS-only and RPS with tidal interaction configuration, hinting towards a very weak (or even negligible) tidal interaction occurring in the past in this system. 
The significance of this claim is highlighted by the fact that we were forced to exclude an even stronger tidal interaction (closer passage between the two galaxy models than fiducial 50 kpcs) because of the appearance of features as strong barring in the NGC 2276 model. 
In addition, the model with an initial separation greater than 50 kpcs would likely show no signature on the gas or stellar properties of our model NGC 2276 galaxy.

The underlying reason why this galaxy has such a high SFR burst in the last 30 Myrs needs to be investigated further.
In our follow-up paper, we aim to investigate the detailed properties of ionized gas, which should bring some light on the origin of this SFR excess.

\begin{acknowledgements}
We would like to thank the anonymous referee for the valuable suggestions and comments which significantly improved this work.
This work was supported by the Croatian Science Foundation under the project number [HRZZ-MOBDOK-2023-8006].
This research has made use of the NASA/IPAC Infrared Science Archive, which is funded by the National Aeronautics and Space Administration and operated by the California Institute of Technology.  
This research is based on observations made with the Galaxy Evolution Explorer, obtained from the MAST data archive at the Space Telescope Science Institute, which is operated by the Association of Universities for Research in Astronomy, Inc., under NASA contract NAS 5–26555.
This publication makes use of data products from the Two Micron All Sky Survey, which is a joint project of the University of Massachusetts and the Infrared Processing and Analysis Center/California Institute of Technology, funded by the National Aeronautics and Space Administration and the National Science Foundation.
This work is based in part on archival data obtained with the Spitzer Space Telescope, which was operated by the Jet Propulsion Laboratory, California Institute of Technology under a contract with NASA.
This work is also based on observations collected at the Centro Astronómico Hispano-Alemán (CAHA), operated jointly by the Max-Planck Institut für Astronomie and the Instituto de Astrofísica de Andalucía (CSIC). 
AI acknowledges support from the institutional project RVO:67985815 and the project 25-19512L of the Czech Science Foundation.
R.S. acknowledges financial support from FONDECYT Regular 2023 project No. 1230441 and also gratefully acknowledges financial support from ANID - MILENIO NCN2024\_112. 
A.E.L. acknowledges support from the INAF GO grant 2023 ''Identifying ram pressure induced unwinding arms in cluster spirals'' (P.I. B. Vulcani).
P.J.W. acknowledges support from the European Union – NextGenerationEU RFF M4C2 1.1 PRIN 2022 project 2022ZSL4BL INSIGHT, the INAF Large Grant 2022 ''Extragalactic Surveys with JWST'' (PI Pentericci), and the INAF Mini Grant ''1.05.24.07.01 RSN1: Spatially Resolved Near-IR Emission of Intermediate-Redshift Jellyfish Galaxies'' (PI Watson).

\end{acknowledgements}

\bibliographystyle{aa}
\bibliography{bibliography}

\begin{appendix}

\section{Optical IFU data calibration}
\label{sect:app-ifu}
The optical IFU observations were carried out with the PMAS instrument \citep{Roth05} in the PPaK mode \citep{Kelz06}, on the Calar Alto 3.5 m telescope in Spain (PI: S. Meidt; project no. F16-3.5-021).  
The observations cover the entire galactic disk with seven pointings ($\approx 72''$ in diameter per pointing) with three dither patterns (with $\Delta [RA, DEC]=[0'', 0'']$, $\Delta [RA, DEC]=[-5.22'', -4.84'']$ and $\Delta [RA, DEC]=[-5.22'', +4.84'']$ offsets), and with V500 grating.
The raw data have been calibrated using the standard procedures that are described in \citet{Tomicic17}, with sky spectra observed from the IFU sky fibers located outside the galactic disk during the observations.
Astrometry and flux calibration have been implemented using photometric maps from the WFC3/HST images that are degraded to the same spatial resolution and pixel grid as the IFU cube, as was done in M26.  
The final mosaic IFU cube covers the entire galactic disk, has spatial resolution of $2.7''$ (and spaxel size of $1''$), and the mean spectral resolution of $R\approx850$.

\section{Sectors and radial bining scheme}

\begin{figure}[ht!]
    \centering
    \includegraphics[width = 0.99\hsize]{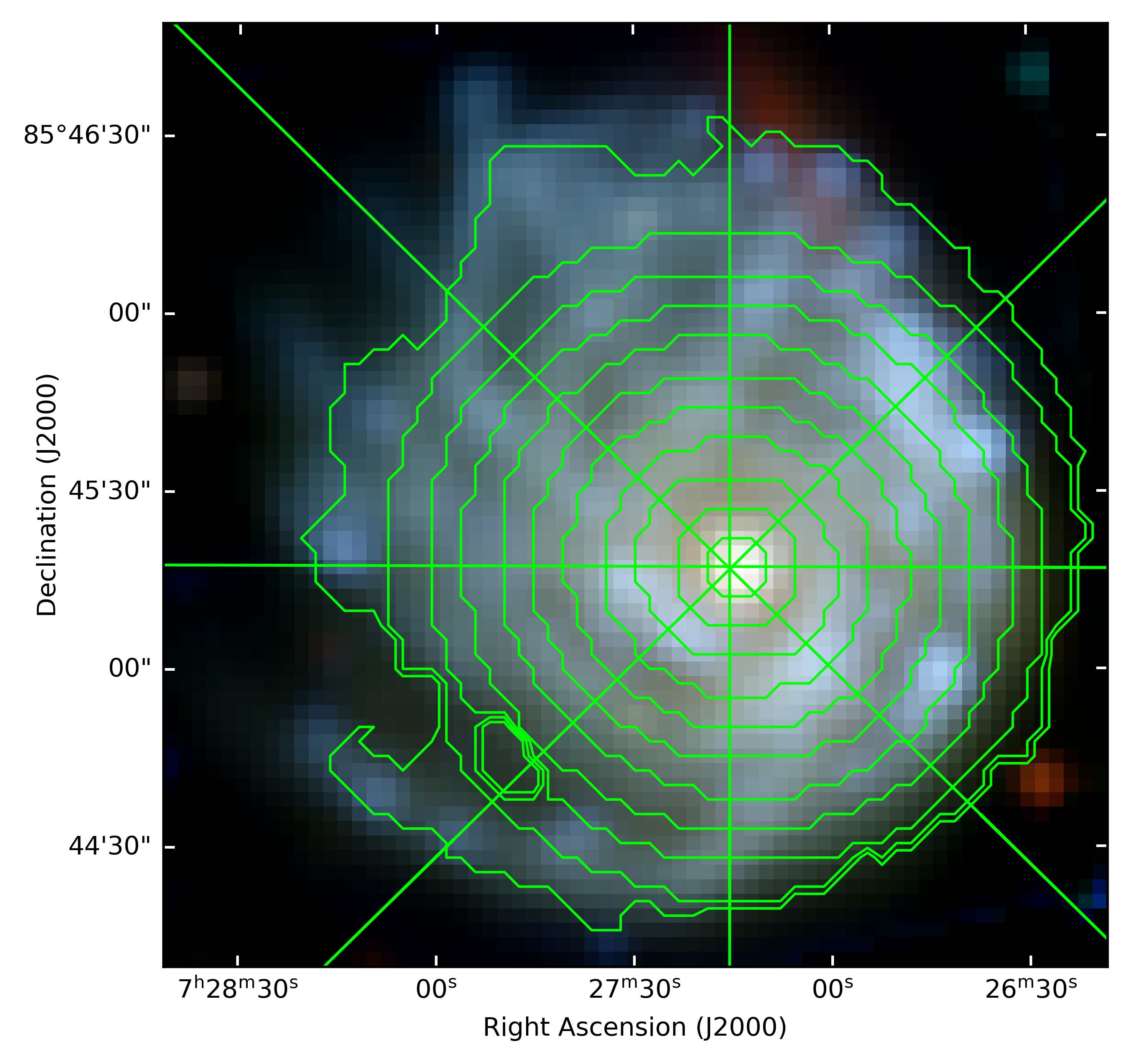}
    \caption{RGB image of NGC 2276 with the radial binning scheme introduced in Sect. \ref{sect:results}, and sectors introduced in M26 overplotted as green contours.
    }
    \label{fig:binning}
\end{figure}

\section{Ancillary \textsc{bagpipes} output maps}

\begin{figure*}[bt!]
    \centering
    \includegraphics[width = 0.99\hsize]{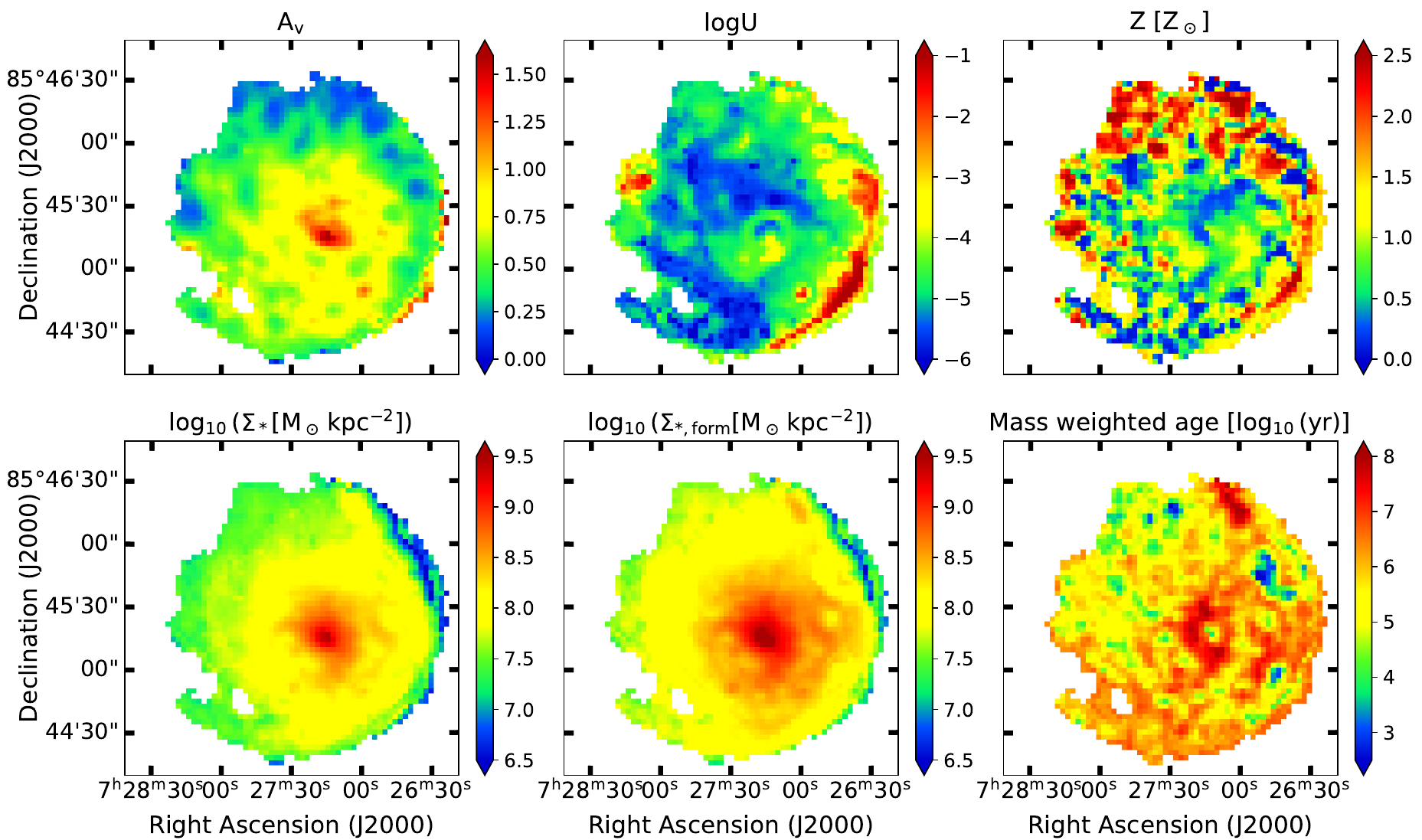}
    \caption{Additional outputs of \textsc{bagpipes}. Top: Attenuation map, ionization parameter map, and metallicity map. Bottom: Stellar mass map, formed stellar mass map, and mass weighted age map. All maps are delimited by the 3-$\sigma$ level contour from the MIPS24 map.
    }
    \label{fig:bagpipes_ancillary}
\end{figure*}

In Fig. \ref{fig:bagpipes_ancillary} we show additional maps derived with \textsc{bagpipes} that were not the main focus of this work. However, the significance of these maps is important for future research of this group.
Due to the limited observational constraints on logU and metallicity, we present these parameters as derived from our SED fitting, and stress that the uncertainties on these maps prevent any robust physical interpretation.

From the stellar A$_\mathrm{V}$ map shown in top-left corner we can see that the attenuation is lowest in the northern and easternmost points of the galaxy, while the regions of the highest attenuation correlate well with the $^{12}$CO($J$=1$\rightarrow$0) map shown in \citet{tomicic18}.

The map of the ionization parameter $U$ shown on the top-center panel indicates that the areas of highest ionization are located mostly on the impact front with the IGM.

Stellar metallicity map shown on the top-right panel shows a scattered distribution. 
Considering that the SED fitting was done with photometric maps and narrow-band synthetic filters around the emission lines, these results have significant uncertainties.

The stellar mass map (shown on bottom-left panel), and the formed stellar mass map (shown on bottom-center panel) indicate that the distribution of the stellar mass is mostly symmetrical. 
Additionally, as suggested from the SFH maps shown on Fig. \ref{fig:sfh}, there is an extended distribution of stellar mass on the eastern side of the galaxy.

Finally, the mass weighted age map shown on the bottom-right panel shows the distribution of the average age of a stellar population, calculated by weighting each star's age by its mass. The areas with the youngest stars seem to be located around the ULX sources discovered by \citet{wolter+15} and \citet{Anastasopoulou+19}.

\section{Additional examples of \textsc{bagpipes} outputs}

\begin{figure*}[b!] 
    \centering
    \begin{tikzpicture}
  
        \node[anchor=south west, inner sep=0pt] (base_img) at (0,0) {
            \includegraphics[width=0.995\textwidth]{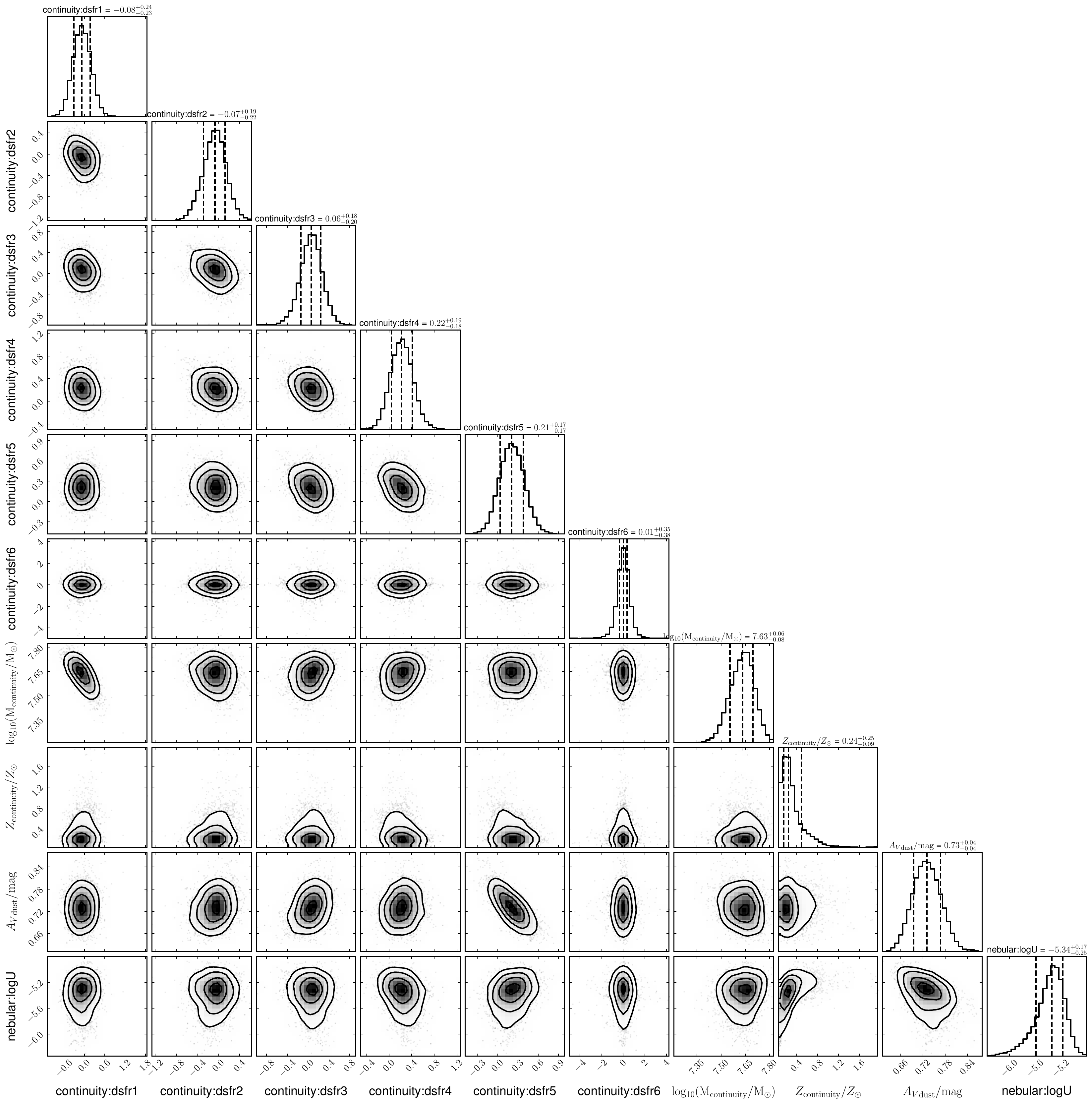}
        };

        \node[anchor=north west, inner sep=0pt] at (11.2cm, 14.25cm) {
            \includegraphics[width=7cm]{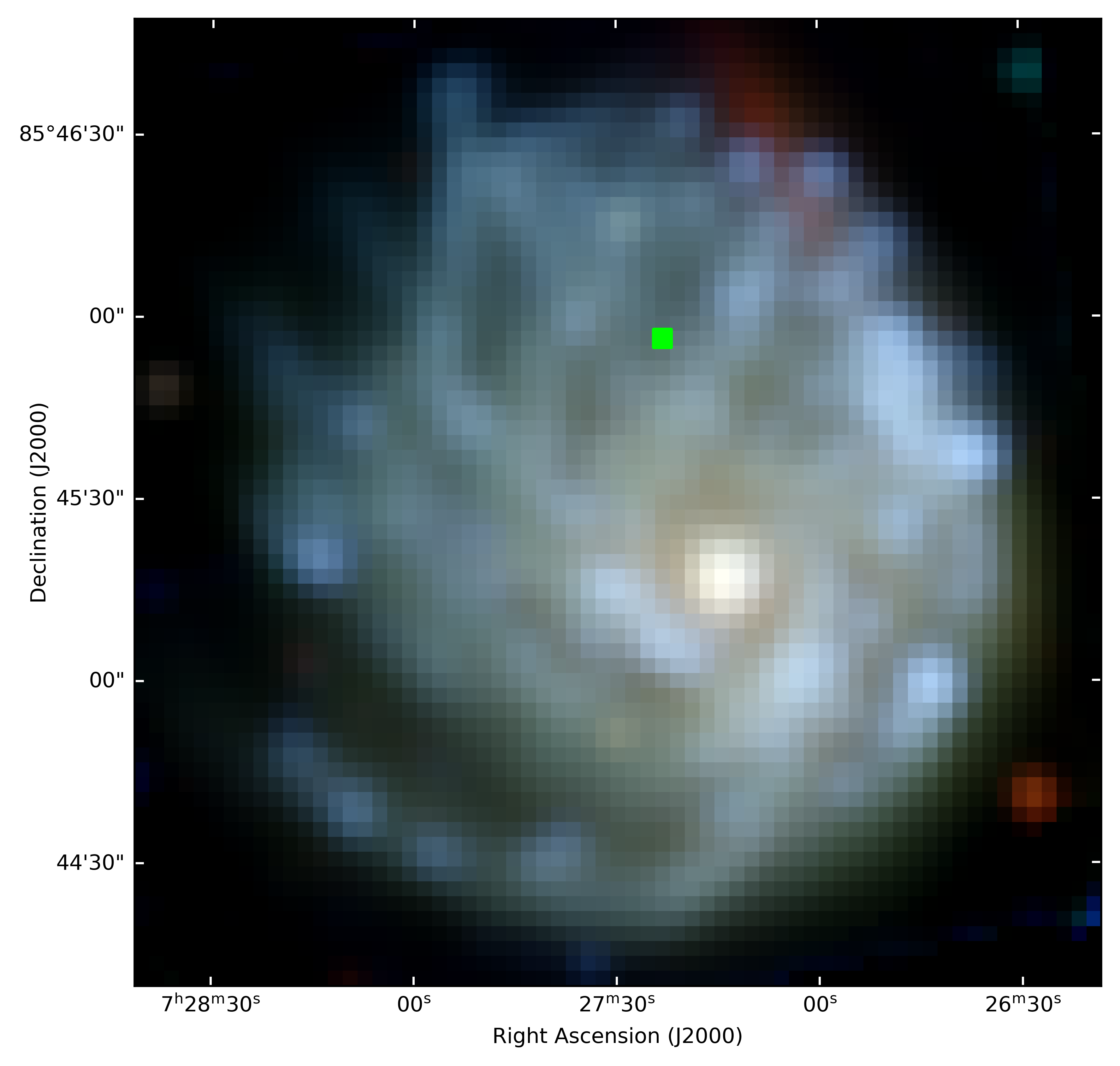}
        };

        \node[anchor=north west, inner sep=0pt] at (6.2cm, 18.5cm) {
            \includegraphics[width=12.2cm]{figs/36+44_fit.pdf}
        };

    \end{tikzpicture}
    \caption{Example of \textsc{bagpipes} output for a selected spatial bin from the fully resolved case in the inter-arm region of NGC 2276, shown with green square on the RGB image. Photometric data is shown on the top right panel in blue (the bands used for fitting are shown in Table \ref{fig:filters}). The 16–84th percentile range for the posterior spectrum and photometry is shaded orange. A corner plot showing the posterior distribution for fitted parameters is shown below. Fitted spectrum and RGB image were already shown on Fig. \ref{fig:fit_ex}.}
    \label{fig:fit_ex1}
\end{figure*}

\begin{figure*}[b!] 
    \centering
    \begin{tikzpicture}

        \node[anchor=south west, inner sep=0pt] (base_img) at (0,0) {
            \includegraphics[width=0.995\textwidth]{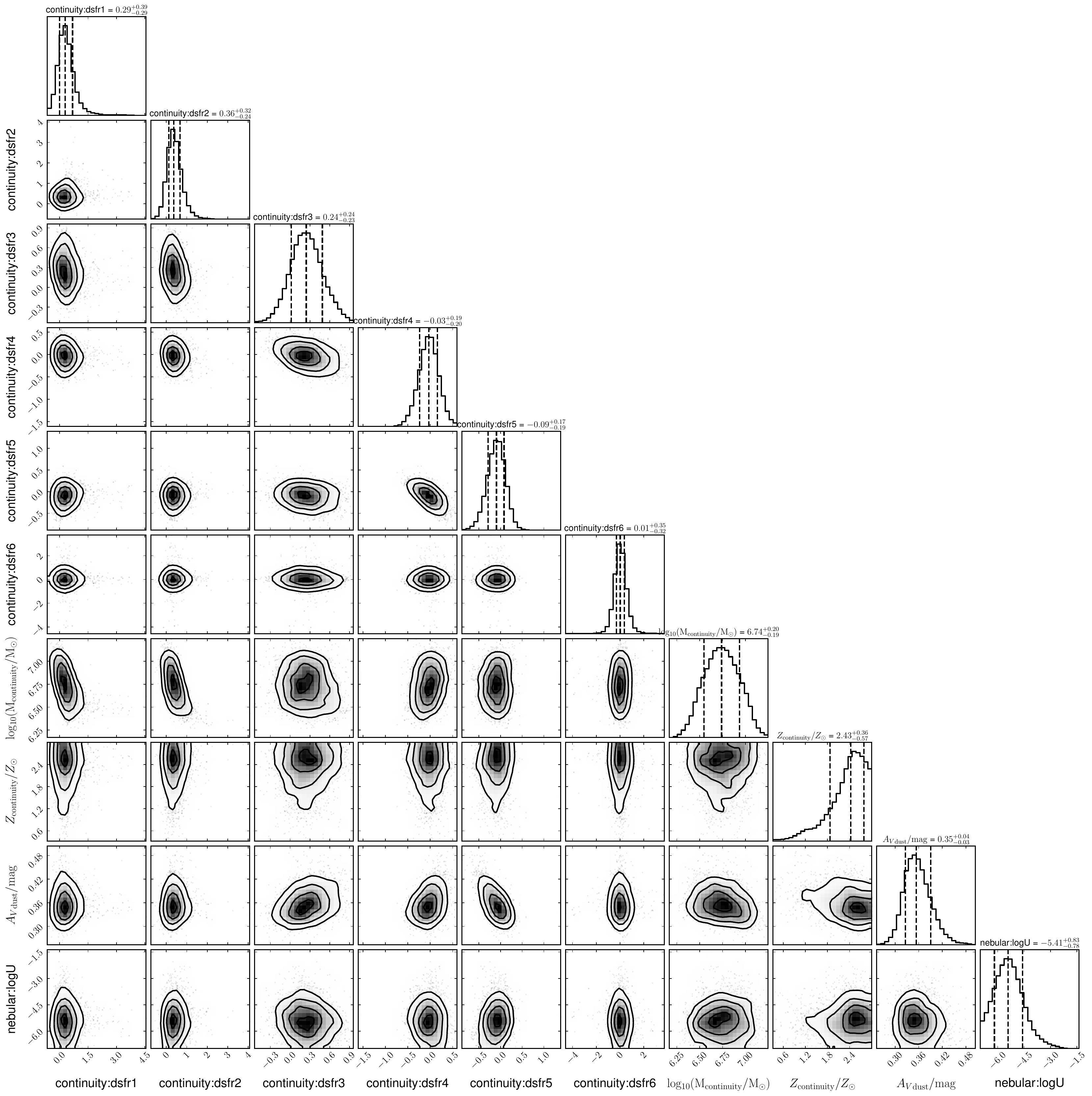}
        };

        \node[anchor=north west, inner sep=0pt] at (11.2cm, 14.25cm) {
            \includegraphics[width=7cm]{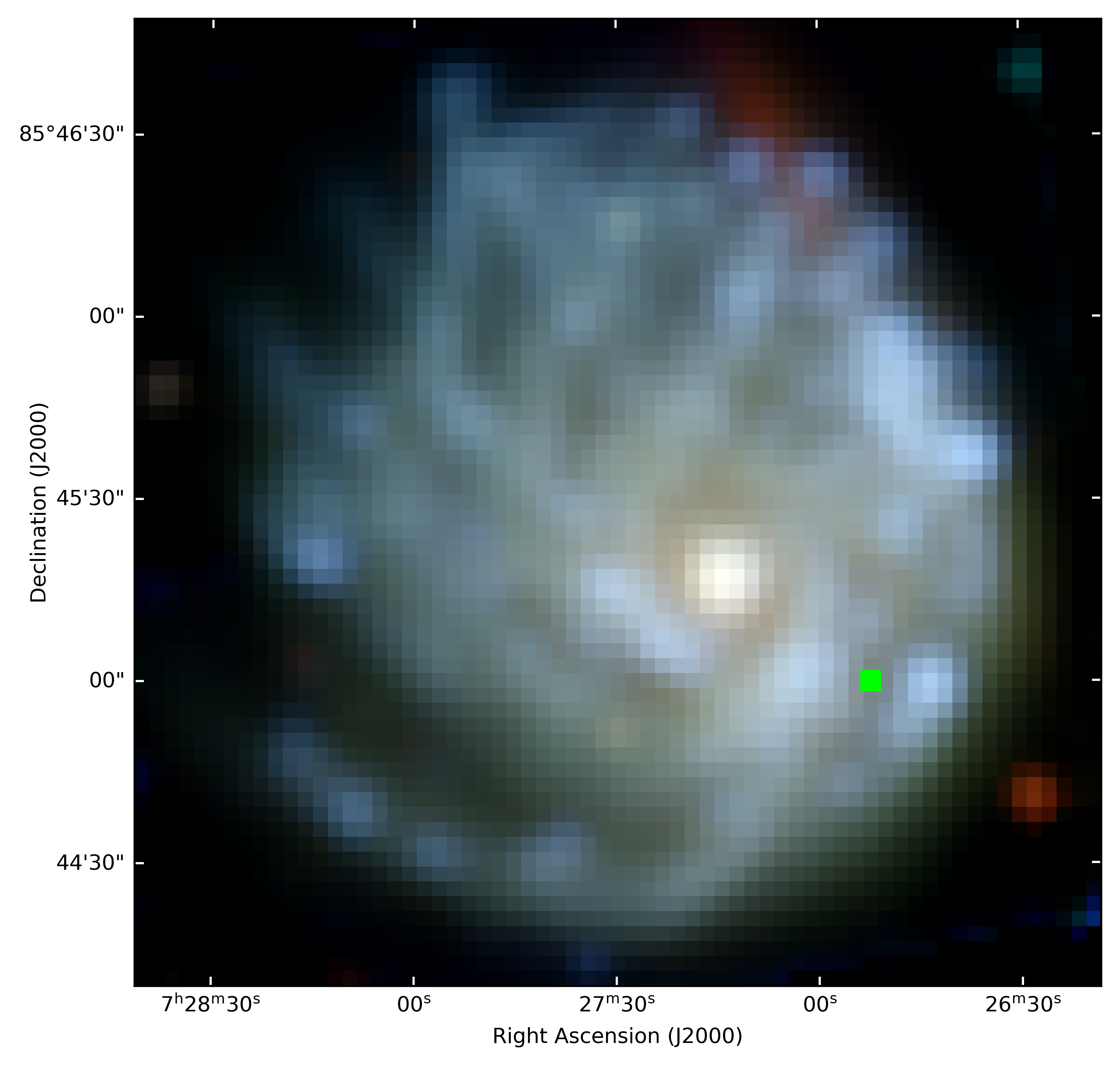}
        };

        \node[anchor=north west, inner sep=0pt] at (6.2cm, 18.5cm) {
            \includegraphics[width=12.2cm]{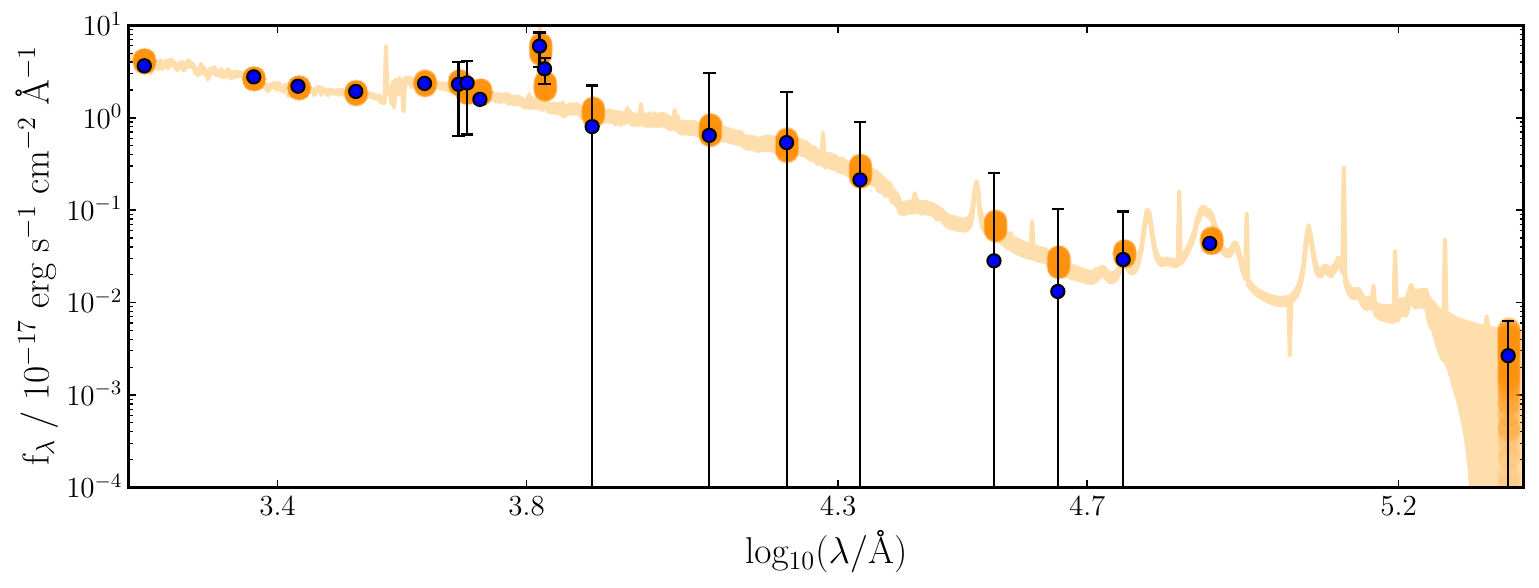}
        };

    \end{tikzpicture}
    \caption{Example of \textsc{bagpipes} output for a selected spatial bin from the fully resolved case on the leading side of NGC 2276, shown with green square on the RGB image. Photometric data is shown on the top right panel in blue (the bands used for fitting are shown in Table \ref{fig:filters}). The 16–84th percentile range for the posterior spectrum and photometry is shaded orange. A corner plot showing the posterior distribution for fitted parameters is shown below.}
    \label{fig:fit_ex3}
\end{figure*}

\begin{figure*}[b!] 
    \centering
    \begin{tikzpicture}

        \node[anchor=south west, inner sep=0pt] (base_img) at (0,0) {
            \includegraphics[width=0.995\textwidth]{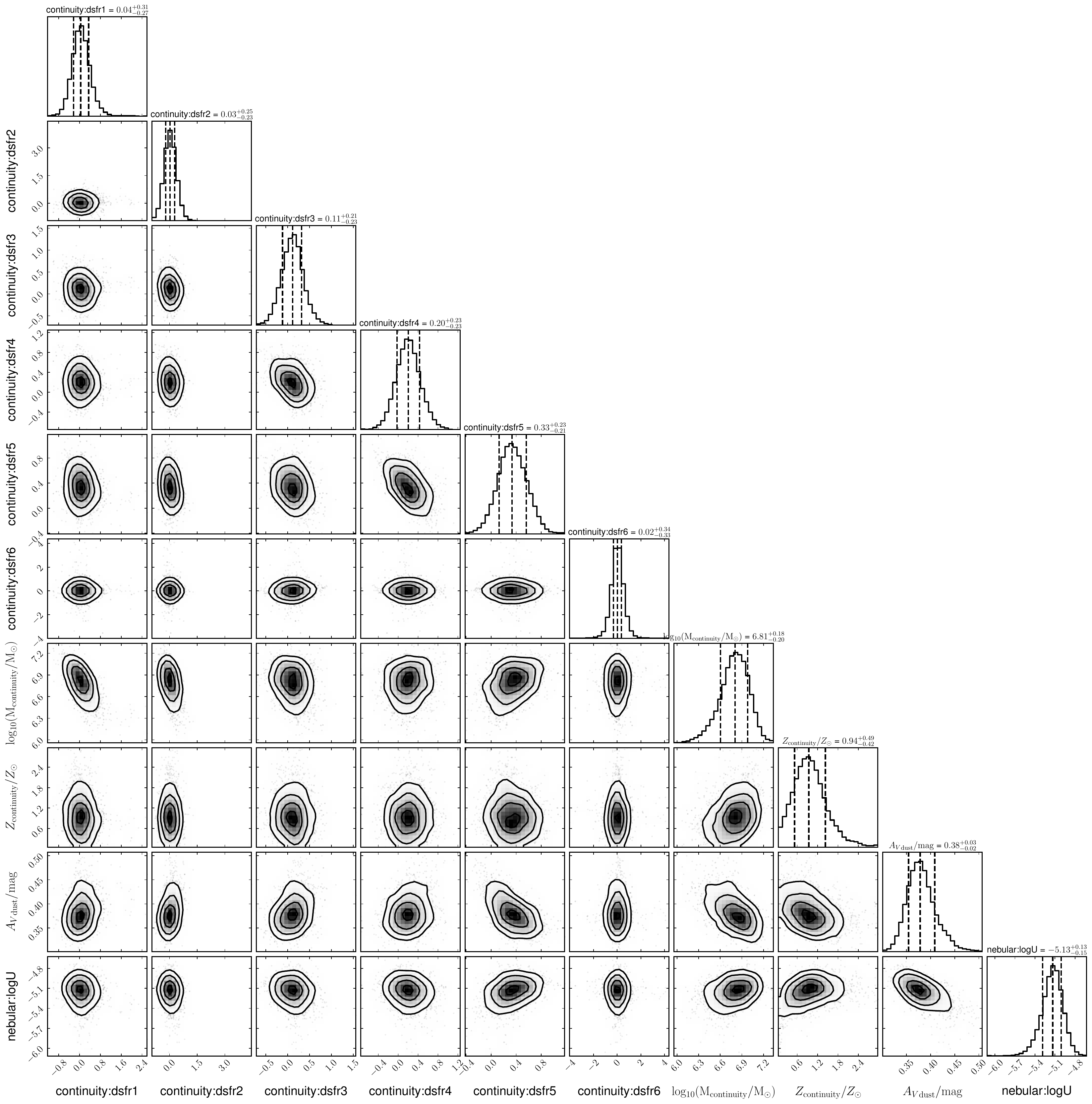}
        };

        \node[anchor=north west, inner sep=0pt] at (11.2cm, 14.25cm) {
            \includegraphics[width=7cm]{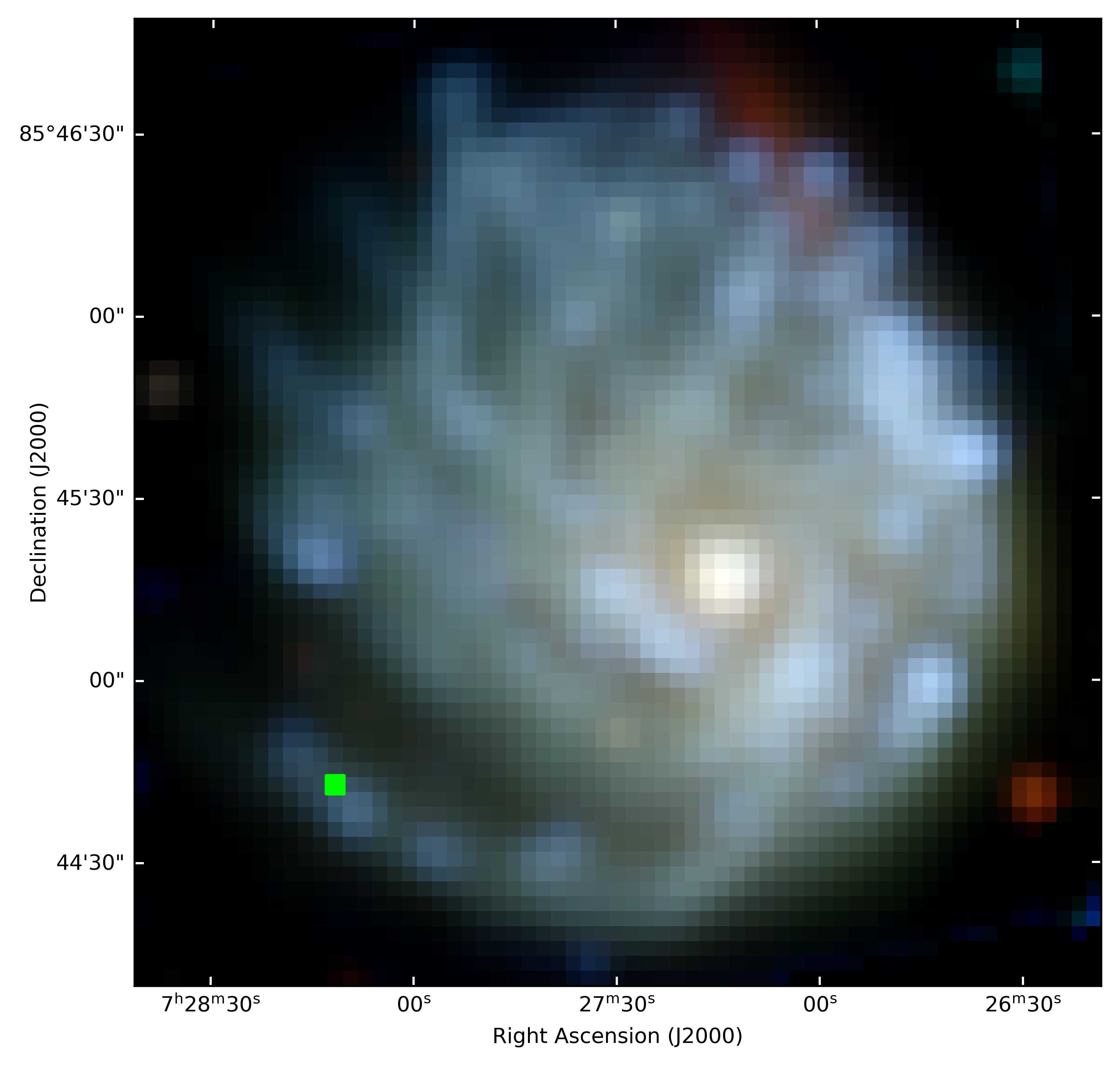}
        };

        \node[anchor=north west, inner sep=0pt] at (6.2cm, 18.5cm) {
            \includegraphics[width=12.2cm]{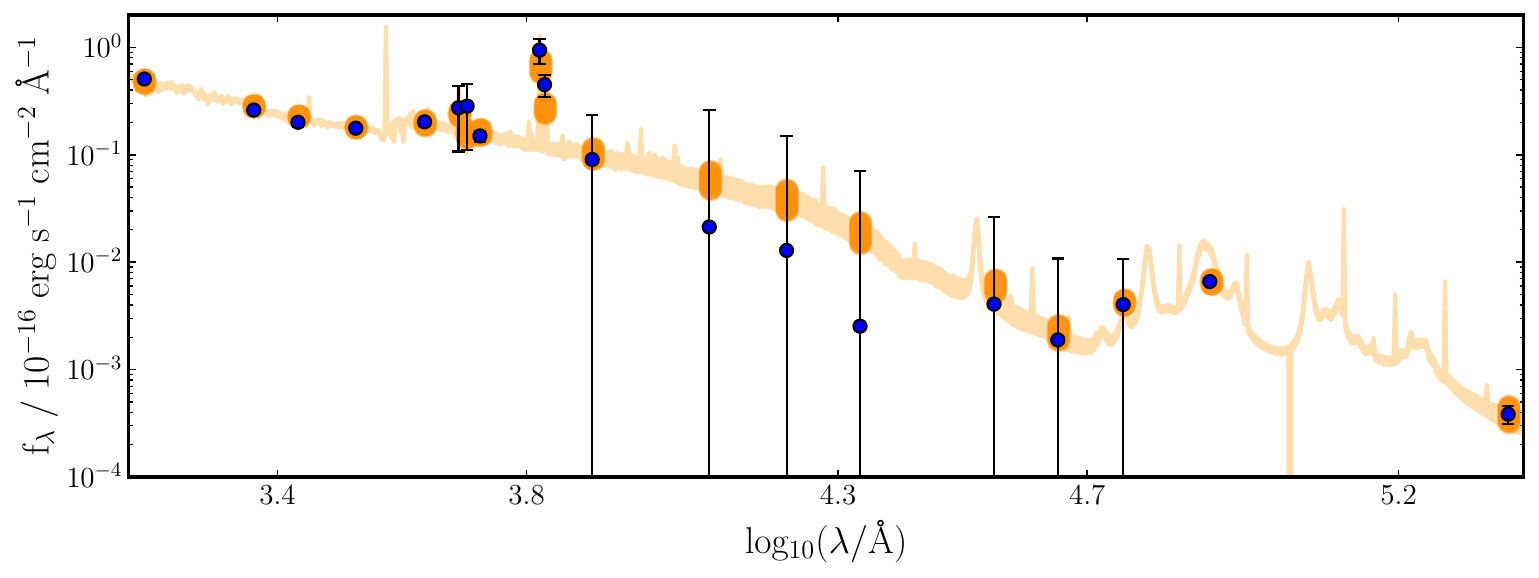}
        };

    \end{tikzpicture}
    \caption{Example of \textsc{bagpipes} output for a selected spatial bin from the fully resolved case in one of the spiral arms of NGC 2276, shown with green square on the RGB image. Photometric data is shown on the top right panel in blue (the bands used for fitting are shown in Table \ref{fig:filters}). The 16–84th percentile range for the posterior spectrum and photometry is shaded orange. A corner plot showing the posterior distribution for fitted parameters is shown below.}
    \label{fig:fit_ex4}
\end{figure*}

In Fig. \ref{fig:fit_ex1}, \ref{fig:fit_ex3}, and \ref{fig:fit_ex4} we show \textsc{bagpipes} SED fits and corner plots for three spaxels in the galactic disk.

\end{appendix}

\end{document}